\documentclass[12pt]{article}
\usepackage{amsfonts}
\usepackage{amsmath}
\usepackage{amsthm}
\usepackage{multirow}
\usepackage{graphicx}
\usepackage{epstopdf}
\usepackage{caption}
\usepackage{subcaption}
\usepackage[right=2.1cm,left=2.1cm, top=2.5cm, bottom=2.5cm]{geometry}

\title{Sampling latent states for high-dimensional non-linear \\ state space models with the embedded HMM method}
\author{Alexander Y. Shestopaloff \\
Department of Statistical Sciences \\
University of Toronto \\
alexander@utstat.utoronto.ca \\
\and Radford M. Neal \\
Department of Statistical Sciences \\
\& Department of Computer Science \\
University of Toronto \\
radford@utstat.utoronto.ca}
\date{18 February 2016, revised 25 June 2016}
\begin{document}

\maketitle

\begin{abstract}
We propose a new scheme for selecting pool states for the embedded Hidden Markov Model (HMM) Markov Chain Monte Carlo (MCMC) method. This new scheme allows the embedded HMM method to be used for efficient sampling in state space models where the state can be high-dimensional. Previously, embedded HMM methods were only applied to models with a one-dimensional state space. We demonstrate that using our proposed pool state selection scheme, an embedded HMM sampler can have similar performance to a well-tuned sampler that uses a combination of Particle Gibbs with Backward Sampling (PGBS) and Metropolis updates. The scaling to higher dimensions is made possible by selecting pool states locally near the current value of the state sequence. The proposed pool state selection scheme also allows each iteration of the embedded HMM sampler to take time linear in the number of the pool states, as opposed to quadratic as in the original embedded HMM sampler. We also consider a model with a multimodal posterior, and show how a technique we term ``mirroring'' can be used to efficiently move between the modes.

\end{abstract}

\section{Introduction}

Consider a non-linear, non-Gaussian state space model for an observed sequence  $y = (y_{1}, \ldots, y_{n})$. This model, with parameters $\theta$, assumes that the $Y_{i}$ are drawn from an observation density $p(y_{i} | x_{i}, \theta)$, where $X_{i}$ is an unobserved Markov process with initial density $p(x_{1}|\theta)$ and transition density $p(x_{i}|x_{i-1}, \theta)$. Here, the $x_{i}$ might be either continuous or discrete. We may be interested in inferring both the realized values of the Markov process $x = (x_{1}, \ldots, x_{n})$ and the model parameters $\theta$. In a Bayesian approach to this problem, this can be done by drawing a sample of values for $x$ and $\theta$ using a Markov chain that alternately samples from the conditional posterior distributions $p(x|\theta, y)$ and $p(\theta|x, y)$. In this paper, we will only consider inference for $x$ by sampling from $p(x|\theta, y)$, taking the parameters $\theta$ to be known. As a result, we will omit $\theta$ in model densities for the rest of the paper. Except for linear Gaussian models and models with a finite state space, this sampling problem has no exact solution and hence approximate methods such as MCMC must be used.

One method for sampling state sequences in non-linear, non-Gaussian state space models is the embedded HMM method (Neal, 2003; Neal, Beal and Roweis, 2004). An embedded HMM update proceeds as follows. First, at each time $i$, a set of $L$ ``pool states'' in the latent space is constructed. In this set, $L-1$ of the pool states are drawn from a chosen pool state density and one is the current value of $x_{i}$. This step can be thought of as temporarily reducing the state space model to an HMM with a finite set of $L$ states, hence the name of the method. Then, using efficient forward-backward computations, which take time proportional to $L^{2}n$, a new sequence $x'$ is selected from the ``ensemble'' of $L^{n}$ sequences passing through the set of pool states, with the probability of choosing each sequence proportional to its posterior density divided by the probability of the sequence under the pool state density. At the next iteration of the sampler, a new set of pool states is constructed, so that the chain can sample all possible $x_{i}$, even when the set of possible values is infinite.

Another method is the Particle Gibbs with Backward Sampling (PGBS) method. The Particle Gibbs (PG) method was first introduced in Andrieu, Doucet and Holenstein (2010); Whiteley suggested the backward sampling modification in the discussion following this paper. Lindsten and Schon (2012) implemented backward sampling and showed that it improves the efficiency of PG. Starting with a current sequence $x$, PGBS first uses conditional Sequential Monte Carlo (SMC) to construct a set of candidate sequences and then uses backward sampling to select a new sequence from the set of candidate ones. Here, conditional SMC works in the same way as ordinary SMC when generating a set of particles, except that one of the particles at time $i$ is always set to the current $x_{i}$, similar to what is done in the embedded HMM method, which allows the sampler to remain at $x_{i}$ if $x_{i}$ lies in a high-density region. While this method works well for problems with low-dimensional state spaces, the reliance of the SMC procedure on choosing an appropriate importance density can make it challenging to make the method work in high dimensions. An important advantage of Particle Gibbs, however, is that each iteration takes time that is only linear in the number of particles.

Both the PGBS and embedded HMM methods can facilitate sampling of a latent state sequence, $x$, when there are strong temporal dependencies amongst the $x_{i}$. In this case, using a  method that samples $x_{i}$ conditional on fixed values of $x_{i-1}$ and $x_{i+1}$ can be an inefficient way of producing a sample from $p(x|y, \theta)$, because the conditional density of $x_{i}$ given $x_{i-1}$ and $x_{i+1}$ can be highly concentrated relative to the marginal density of $x_{i}$. In contrast, with the embedded HMM and PGBS methods it is possible to make changes to blocks of $x_{i}$'s at once. This allows larger changes to the state in each iteration of the sampler, making updates more efficient. However, good performance of the embedded HMM and PGBS methods relies on appropriately choosing the set of pool states or particles at each time $i$.

In this paper, our focus will be on techniques for choosing pool states for the embedded HMM method. When the latent state space is one-dimensional, embedded HMMs work well when choosing pool states in a variety of ways. For example, in Shestopaloff and Neal (2013), we choose pool states at each time $i$ by constructing a ``pseudo-posterior'' for each latent variable by taking the product of a ``pseudo-prior'' and the observation density, the latter treated as a ``pseudo-likelihood'' for the latent variable. In Shestopaloff and Neal (2014), we choose pool states at each time $i$ by sampling from the marginal prior density of the latent process.

Ways of choosing pool states that work well in one dimension begin to exhibit problems when applied to models with higher-dimensional state spaces. This is true even for dimensions as small as three. Since these schemes are global, designed to produce sets of pool states without reference to the current point, as the dimension of the latent space grows, a higher proportion of the sequences in the ensemble ends up having low posterior density. Ensuring that performance doesn't degrade in higher dimensions thus requires a significant increase in the number of pool states. As a result, computation time may grow so large that any advantage that comes from using embedded HMMs is eliminated. One advantage of the embedded HMM method over PGBS is that the embedded HMM construction allows placing pool states locally near the current value of $x_{i}$, potentially allowing the method to scale better with the dimensionality of the state space. Switching to such a local scheme fixes the problem to some extent. However, local pool state schemes come with their own problems, such as making it difficult to handle models with multiple posterior modes that are well-separated --- the pool states might end up being placed near only some of the modes.

In this paper, we propose an embedded HMM sampler suitable for models where the state space is high dimensional. This sampler uses a sequential approximation to the density $p(x_{i} | y_{1}, \ldots y_{i})$ or to the density $p(x_{i}|y_{i+1}, \ldots, y_{n})$ as the pool state density. We show that by using this pool state density, together with an efficient MCMC scheme for sampling from it, we can reduce the cost per iteration of the embedded HMM sampler to be proportional to $nL$, as with PGBS. At the same time, we retain the ability to generate pool states locally, allowing better scaling for high-dimensional state spaces. Our proposed scheme can thus be thought of as combining the best features of the PGBS and the embedded HMM methods, while overcoming the deficiencies of both. We use two sample state space models as examples. Both have Gaussian latent processes and Poisson observations, with one model having a unimodal posterior and the second a multimodal one. For the multimodal example, we introduce a ``mirroring'' technique that allows efficient movement between the different posterior modes. For these models, we show how our proposed embedded HMM method compares to a simple Metropolis sampler, a PGBS sampler, as well as a sampler that combines PGBS and simple Metropolis updates. Further details on ensemble methods are available in the PhD thesis of Shestopaloff (2016).

\section{Embedded HMM MCMC}

We review the embedded HMM method (Neal, 2003; Neal, Beal and Roweis, 2004) here. We take the model parameters, $\theta$, to be fixed, so we do not write them explicitly. Let $p(x)$ be the density from which the state at time $1$ is drawn, let $p(x_{i}|x_{i-1})$ be the transition density between states at times $i$ and $i-1$, and let $p(y_{i}|x_{i})$ be the density of the observation $y_{i}$ given $x_{i}$. 

Suppose our current sequence is $x = (x_{1}, \ldots, x_{n})$. The embedded HMM sampler updates $x$ to $x'$ as follows.

First, at each time $i = 1, \ldots, n$, we generate a set of $L$ pool states, denoted by $\mathcal{P}_{i} = \{x_{i}^{[1]}, \ldots, x_{i}^{[L]}\}$. The pool states are sampled independently across the different times $i$. We choose $l_{i} \in \{1, \ldots, L\}$ uniformly at random and set $x_{i}^{[l_{i}]}$ to $x_{i}$. We sample the remaining $L-1$ pool states $x_{i}^{[1]}, \ldots, x_{i}^{[l_{i}-1]}, x_{i}^{[l_{i}+1]}, \ldots, x_{n}^{[L]}$ using a Markov chain that leaves a pool density $\kappa_{i}$ invariant, as follows. Let $R_{i}(x'|x)$ be the transitions of this Markov chain with $\tilde{R}_{i}(x|x')$ the transitions for this Markov chain reversed (i.e.\ $\tilde{R_{i}}(x|x') = R_{i}(x'|x)\kappa_{i}(x)/\kappa_{i}(x)$), so that
\begin{eqnarray}
\kappa_{i}(x)R_{i}(x'|x) = \kappa_{i}(x')\tilde{R_{i}}(x|x')
\end{eqnarray}
for all $x$ and $x'$. Then, starting at $j = l_{i} - 1$, use reverse transitions $\tilde{R}_{i}(x_{i}^{[j]}|x_{i}^{[j+1]})$ to generate $x_{i}^{[l_{i}-1]}, \ldots, x_{i}^{[1]}$ and starting at $j = l_{i} + 1$ use forward transitions $R_{i}(x_{i}^{[j]}|x_{i}^{[j-1]})$ to generate $x_{i}^{[l_{i}+1]}, \ldots, x_{n}^{[L]}$.

At each $i = 1, \ldots, n$, we then compute the forward probabilities $\alpha_{i}(x)$, with $x$ taking values in $\mathcal{P}_{i}$. At time $i = 1$, we have
\begin{eqnarray}
\alpha_{1}(x) =\frac{p(x)p(y_{1}|x)}{\kappa_{1}(x)}
\label{eq:alphainit}
\end{eqnarray}
and at times $i = 2, \ldots, n$, we have
\begin{eqnarray}
\alpha_{i}(x) = \frac{p(y_{i}|x)}{\kappa_{i}(x)}\sum_{l=1}^{L}p(x|x_{i-1}^{[l]})\alpha_{i-1}(x_{i-1}^{[l]})
\label{eq:alphaall}
\end{eqnarray}

Finally, we sample a new state sequence $x'$ using a stochastic backwards pass. This is done by selecting $x_{n}'$ amongst the set, $\mathcal{P}_{n}$, of pool states at time $n$, with probabilities proportional to $\alpha_{n}(x)$, and then going backwards, sampling $x_{i-1}'$ from the set $P_{i-1}$, with probabilities proportional to $\alpha_{i-1}(x)p(x_{i}'|x)$. Note that only the relative values of the $\alpha_{i}(x)$ will be required, so the $\alpha_{i}$ may be computed up to some constant factor.

Alternatively, given a set of pool states, embedded HMM updates can be done by first computing the backward probabilities. We will see later on that the backward probability formulation of the embedded HMM method allows us to introduce a variation of our proposed pool state selection scheme. Setting $\beta_{n}(x) = 1$ for all $x \in \mathcal{P}_{n}$, we compute for $i < n$
\begin{eqnarray}
\beta_{i}(x) = \frac{1}{\kappa_{i}(x)}\sum_{l=1}^{L}p(y_{i+1}|x_{i+1}^{[l]})p(x_{i+1}^{[l]}|x)\beta_{i+1}(x_{i+1}^{[l]})
\label{eq:beta}
\end{eqnarray}

A new state sequence is then sampled using a stochastic forward pass, setting $x_{1}'$ to one of the $x$ in the pool $\mathcal{P}_{1}$ with probabilities proportional to $\beta_{1}(x)p(x)p(y_{1}|x)$ and then choosing subsequent states $x_{i}'$ from the pools $\mathcal{P}_{i}$ with probabilities proportional to $\beta_{i}(x)p(x|x_{i-1}')p(y_{i}|x)$.

Computing the $\alpha_{i}$ or $\beta_{i}$ at each time $i > 1$ takes time proportional to $L^{2}$, since for each of the $L$ pool states it takes time proportional to $L$ to compute the sums in (\ref{eq:alphaall}) or (\ref{eq:beta}). Hence each iteration of the embedded HMM sampler takes time proportional to $L^{2}n$.

\section{Particle Gibbs with Backward Sampling MCMC}

We review the Particle Gibbs with Backward Sampling (PGBS) sampler here. For full details, see the articles by Andrieu, Doucet and Holenstein (2010) and Lindsten and Schon (2012).

Let $q_{1}(x|y_{1})$ be the importance density from which we sample particles at time $1$, and let $q_{i}(x|y_{i}, x_{i-1})$ be the importance density for sampling particles at times $i > 1$. These may depend on the current value of the parameters, $\theta$, which we suppressed in this notation. Suppose we start with a current sequence $x$. We set the first particle $x_{1}^{[1]}$ to the current state $x_{1}$. We then sample $L-1$ particles $x_{1}^{[2]}, \ldots, x_{1}^{[L]}$ from $q_{1}$  and compute and normalize the weights of the particles:
\begin{eqnarray}
w_{1}^{[l]} &=& \frac{p(x_{1}^{[l]})p(y_{1}|x_{1}^{[l]})}{q_{1}(x_{1}^{[l]}|y_{1})} \\
W_{1}^{[l]} &=& \frac{w_{1}^{[l]}}{\sum_{m=1}^{L}w_{1}^{[m]}}
\end{eqnarray}
for $l = 1, \ldots, L$.

For $i > 1$, we proceed sequentially. We first set $x_{i}^{[1]} = x_{i}$. We then sample a set of $L-1$ ancestor indices for particles at time $i$, defined by $A_{i-1}^{[l]} \in \{1, \ldots, L\}$, for $l = 2, \ldots, L$, with probabilities proportional to $W_{i-1}^{[l]}$. The ancestor index for the first state, $A_{i-1}^{[1]}$, is $1$. We then sample each of the $L-1$ particles, $x_{i}^{[l]}$, at time $i$, for $l = 2, \ldots, L$, from $q_{i}(x|y_{i}, x_{i-1}^{[A_{i-1}^{[l]}]})$ and compute and normalize the weights at time $i$
\begin{eqnarray}
w_{i}^{[l]} &=& \frac{p(x_{i}^{[l]}|x_{i-1}^{[A_{i-1}^{[l]}]})p(y_{i}|x_{i}^{[l]})}{q_{i}(x_{i}^{[l]}|y_{i},x_{i-1}^{[A_{i-1}^{[l]}]})} \\
W_{i}^{[l]} &=& \frac{w_{i}^{[l]}}{\sum_{m=1}^{L}w_{i}^{[m]}}
\end{eqnarray}
A new sequence taking values in the set of particles at each time is then selected using a backwards sampling pass. This is done by first selecting $x_{n}'$ from the set of particles at time $n$ with probabilities $W_{n}^{[l]}$ and then selecting the rest of the sequence going backward in time to time $1$, setting $x_{i}'$ to $x_{i}^{[l]}$ with probability
\begin{eqnarray}
\frac{w_{i}^{[l]}p(x_{i+1}'|x_{i}^{[l]})}{\sum_{m=1}^{L}w_{i}^{[m]}p(x_{i+1}'|x_{i}^{[m]})}
\end{eqnarray}
A common choice for $q$ is the model's transition density, which is what is compared to in this paper.

Note that each iteration of the PGBS sampler takes time proportional to $Ln$, since it takes time proportional to $L$ to create the set of particles at each time $i$, and to do one step of backward sampling.

\section{An embedded HMM sampler for high dimensions}

We propose two new ways, denoted $f$ and $b$, of generating pool states for the embedded HMM sampler. Unlike previously-used pool state selection schemes, where pool states are selected independently at each time, our new schemes select pool states sequentially, with pool states at time $i$ selected conditional on pool states at time $i-1$, or alternatively at time $i+1$.

\subsection{Pool state distributions}

The first way to generate pool states is to use a forward pool state selection scheme, with a sequential approximation to $p(x_{i}|y_{1}, \ldots, y_{i})$ as the pool state density. In particular, at time $1$, we set the pool state distribution of our proposed embedded HMM sampler to
\begin{eqnarray}
\kappa_{1}^{f}(x) \propto p(x)p(y_{1}|x)
\end{eqnarray}
As a result of equation (\ref{eq:alphainit}), $\alpha_{1}(x)$ is constant. At time $i > 1$, we set the pool state distribution to
\begin{eqnarray}
\kappa_{i}^{f}(x|\mathcal{P}_{i-1}) \propto p(y_{i} | x)\sum_{\ell=1}^{L}p(x|x_{i-1}^{[\ell]})
\end{eqnarray}
which makes $\alpha_{i}(x)$ constant for $i > 1$ as well (see equation (\ref{eq:alphaall})).

We then draw a sequence composed of these pool states with the forward probability implementation of the embedded HMM method, with the $\alpha_{i}(x)$'s all set to $1$.

The second way is to instead use a backward pool state selection scheme, with a sequential approximation of $p(x_{i}|y_{i+1}, \ldots, y_{n})$ as the pool state density. We begin by creating the pool $\mathcal{P}_{n}$, consisting of the current state $x_{n}$ and the remaining $L-1$ pool states sampled from $p_{n}(x)$, the marginal density at time $n$, which is the same as $p(x)$ if the latent process is stationary. The backward probabilities $\beta_{n}(x)$, for $x$ in $\mathcal{P}_{n}$, are then set to $1$. At time $i < n$ we set the pool state densities to
\begin{eqnarray}
\kappa_{i}^{b}(x|\mathcal{P}_{i+1}) \propto \sum_{\ell=1}^{L}p(y_{i+1}|x_{i+1}^{[\ell]})p(x_{i+1}^{[\ell]}|x)
\end{eqnarray}
so that $\beta_{i}(x)$ is constant for all $i = 1, \ldots, n$ (see equation \ref{eq:beta}).

We then draw a sequence composed of these pool states as in the backward probability implementation of the embedded HMM method, with the $\beta_{i}(x)$'s all set to $1$.

If the latent process is Gaussian, and the latent state at time $1$ is  sampled from the stationary distribution of the latent process, it is possible to update the latent variables by applying the forward scheme to the reversed sequence $(y_{n}, \ldots, y_{1})$ by making use of time reversibility, since $X_{n}$ is also sampled from the stationary distribution, and the latent process evolves backward in time according to the same transition density as it would going forward. We then use the forward pool state selection scheme along with a stochastic backward pass to sample a sequence $(x_{n}, \ldots, x_{1})$, starting with $x_{1}$ and going to $x_{n}$.

It can sometimes be advantageous to alternate between using forward and backward (or, alternatively, forward applied to the reversed sequence) embedded HMM updates, since this can improve sampling of certain $x_{i}$. The sequential pool state selection schemes use only part of the observed sequence in generating the pool states. By alternating update directions, the pool states can depend on different parts of the observed data, potentially allowing us to better cover the region where $x_{i}$ has high posterior density. For example, at time $1$, the pool state density may disperse the pool states too widely, leading to poor sampling for $x_{1}$, but sampling $x_{1}$ using a backwards scheme can be much better, since we are now using all of the data in the sequence when sampling pool states at time $1$.

\subsection{Sampling pool states}

To sample from $\kappa_{i}^{f}$ or $\kappa_{i}^{b}$, we can use any Markov transitions $R_{i}$ that leave this distribution invariant. The validity of the method does not depend on the Markov transitions for sampling from $\kappa_{i}^{f}$ or $\kappa_{i}^{b}$ reaching equilibrium or even on them being ergodic.

Directly using these pool state densities in an MCMC routine leads to a computational cost per iteration that is proportional to $L^{2}n$, like in the original embedded HMM method, since at times $i > 1$ we need at least $L$ updates to produce $L$ pool states, and the cost of computing an acceptance probability is proportional to $L$.

However, it is possible to reduce the cost per iteration of the embedded HMM method to be proportional to $nL$ when we use $\kappa_{i}^{f}$ or $\kappa_{i}^{b}$ as the pool state densities. To do this, we start by thinking of the pool state densities at each time $i > 1$ as marginal densities summing over the variable $\ell = 1, \ldots, L$ that indexes a pool state at the previous time. Specifically, $\kappa_{i}^{f}$ can be viewed as a marginal of the density
\begin{eqnarray}
\lambda_{i}(x, \ell) \propto p(y_{i}|x)p(x|x_{i-1}^{[\ell]})
\end{eqnarray}
while $\kappa_{i}^{b}$ is a marginal of the density
\begin{eqnarray}
\gamma_{i}(x, \ell) \propto p(y_{i+1}|x_{i+1}^{[\ell]})p(x_{i+1}^{[\ell]}|x)
\end{eqnarray}
Both of these densities are defined given a pool $\mathcal{P}_{i-1}$ at time $i-1$ or pool $\mathcal{P}_{i+1}$ at time $i+1$. This technique is reminiscent of the auxiliary particle filter of Pitt and Shephard (1999). We then use Markov transitions, $R_{i}$, to sample a set of values of $x$ and $\ell$, with probabilities proportional to $\lambda_{i}$ for the forward scheme, or probabilities proportional to $\gamma_{i}$ for the backward scheme.

The chain is started at $x$ set to the current $x_{i}$, and the initial value of $\ell$ is chosen randomly with probabilities proportional to $p(x_{i} | x_{i-1}^{[\ell]})$ for the forward scheme or $p(y_{i+1}|x_{i+1}^{[\ell]})p(x_{i+1}^{[\ell]}|x_{i})$ for the backward scheme. This stochastic initialization of $\ell$ is needed to make the algorithm valid when we use $\lambda_{i}$ or $\gamma_{i}$ to generate the pool states.

Sampling values of $x$ and $\ell$ from $\lambda_{i}$ or $\gamma_{i}$ can be done by updating each of $x$ and $\ell$ separately, alternately sampling values of $x$ conditional on $\ell$, and values of $\ell$ conditional on $x$, or by updating $x$ and $\ell$ jointly, or by a combination of these.

Updating $x$ given $\ell$ can be done with any appropriate sampler, such as Metropolis, or for a Gaussian latent process we can use autoregressive updates, which we describe below. To update $\ell$ given $x$, we can also use Metropolis updates, proposing $\ell' = \ell + k$, with $k$ drawn from some proposal distribution on $\{-K, \ldots, -1, 1, \ldots, K\}$. Alternatively, we can simply propose $\ell'$ uniformly at random from $\{1, \ldots, L\}$. 

To jointly update $x$ and $\ell$, we propose two novel updates, a ``shift'' update and a ``flip'' update. Since these are also Metropolis updates, using them together with Metropolis or autoregressive updates, for each of $x$ and $\ell$ separately, allows embedded HMM updates to be performed in time proportional to $nL$.

\subsection{Autoregressive updates}

For sampling pool states in our embedded HMM MCMC schemes, as well as for comparison MCMC schemes, we will make use of Neal's (1998) ``autoregressive'' Metropolis-Hastings update, which we review here. This update is designed to draw samples from a distribution of the form $p(w)p(y|w)$ where $p(w)$ is multivariate Gaussian with mean $\mu$ and covariance $\Sigma$ and $p(y|w)$ is typically a density for some observed data.

This autoregressive update proceeds as follows. Let $L$ be the lower triangular Cholesky decomposition of $\Sigma$, so $\Sigma = LL^{T}$, and $n$ be a vector of i.i.d. normal random variables with zero mean and identity covariance. Let $\epsilon \in [-1, 1]$ be a tuning parameter that determines the scale of the proposal. Starting at $w$, we propose
\begin{eqnarray}
w' = \mu + \sqrt{1 - \epsilon^{2}}(w - \mu) + \epsilon L n
\end{eqnarray}
Because these autoregressive proposals are reversible with respect to $p(w)$, the proposal density and $p(w)$ cancel in the Metropolis-Hastings acceptance ratio. This update is therefore accepted with probability 
\begin{eqnarray}
\min\biggl(1, \frac{p(y|w')}{p(y|w)}\biggr)
\end{eqnarray}
Note that for this update, the same value of $\epsilon$ is used for scaling along every dimension. It would be of independent interest to develop a version of this update where $\epsilon$ can be different for each dimension of $w$.

\subsection{Shift updates}

We can simultaneously update $\ell$ and $x$ at time $i > 1$ by proposing to update $(x, \ell)$ to $(x', \ell')$ where $\ell'$ is proposed in any valid way while $x'$ is chosen in a way such that $x'$ and $x_{i-1}^{[\ell']}$ are linked in the same way as $x$ and $x_{i-1}^{[\ell]}$. The shift update makes it easier to generate a set of pool states at time $i$ with different predecessor states at time $i-1$, helping to ensure that the pool states are well-dispersed. This update is accepted with the usual Metropolis probability.

For a concrete example we use later, suppose that the latent process is an autoregressive Gaussian process of order $1$, with the model being that $X_{i} | x_{i-1} \sim N(\Phi x_{i-1}, \Sigma)$. In this case, given $\ell'$, we propose $x_{i}' = x_{i} + \Phi (x_{i-1}^{[\ell']} - x_{i-1}^{[\ell]})$. This update is accepted with probability
\begin{eqnarray}
\min\biggl(1, \frac{p(y_{i}|x_{i}')}{p(y_{i}|x_{i})}\biggr)
\end{eqnarray}
as a result of the transition densities in the acceptance ratio cancelling out, since
\begin{eqnarray}
x_{i}' - \Phi x_{i-1}^{[\ell']} &=& x_{i} + \Phi (x_{i-1}^{[\ell']} - x_{i-1}^{[\ell]}) - \Phi x_{i-1}^{[\ell']} \\
&=& x_{i} - \Phi x_{i-1}^{[\ell]}
\end{eqnarray}

To be useful, shift updates normally need to be combined with other updates for generating pool states. When combining shift updates with other updates, tuning of acceptance rates for both updates needs to be done carefully in order to ensure that the shift updates actually improve sampling performance. In particular, if the pool states at time $i-1$ are spread out too widely, then the shift updates may have a low acceptance rate and not be very useful. Therefore, jointly optimizing proposals for $x$ and for $x$ and $\ell$ may lead to a relatively high acceptance rate on updates of $x$, in order to ensure that the acceptance rate for the shift updates isn't low.

\subsection{Flip updates}

Generating pool states locally can be helpful when applying embedded HMMs to models with high-dimensional state spaces but it also makes sampling difficult if the posterior is multimodal. Consider the case when the observation probability depends on $|x_{i}|$ instead of $x_{i}$, so that many modes with different signs for some $x_{i}$ exist. We propose to handle this problem by adding an additional flip update that creates a ``mirror'' set of pool states, in which $-x_{i}$ will be in the pool if $x_{i}$ is. By having a mirror set of pool states, we are able to flip large segments of the sequence in a single update, allowing efficient exploration of different posterior modes.

To generate a mirror set of pool states, we must correctly use the flip updates when sampling the pool states. Since we want each pool state to have a negative counterpart, we choose the number of pool states $L$ to be even. The chain used to sample pool states then alternates two types of updates, a usual update to generate a pool state and a flip update to generate its negated version. The usual update can be a combination of any updates, such as those we consider above. So that each state will have a flipped version, we start with a flip transition between $x^{[1]}$ and $x^{[2]}$, a usual transition between $x^{[2]}$ and $x^{[3]}$, and so on up to a flip transition between $x^{[L-1]}$ to $x^{[L]}$.

At time $1$, we start with the current state $x_{1}$ and randomly assign it to some index $l_{1}$ in the chain used to generate pool states. Then, starting at $x_{1}$ we generate pool states by reversing the Markov chain transitions back to $1$ and going forward up to $L$. Each flip update is then a Metropolis update proposing to generate a pool state $-x_{1}$ given that the chain is at some pool state $x_{1}$. Note that if the observation probability depends on $x_{1}$ only through $|x_{1}|$ and $p(x)$ is symmetric around zero then this update is always accepted. 

At time $i > 1$, a flip update proposes to update a pool state $(x, \ell)$ to $(-x, \ell')$ such that $x_{i-1}^{[\ell']} = -x_{i-1}^{[\ell]}$. Here, since the pool states at each time are generated by alternating flip and usual updates, starting with a flip update to $x_{i}^{[1]}$, the proposal to move from $\ell$ to $\ell'$ can be viewed as follows. Suppose that instead of labelling our pool states from $1$ to $L$ we instead label them $0$ to $L-1$. The pool states at times $0$ and $1$, then $2$ and $3$, and so on will then be flipped pairs, and the proposal to change $\ell$ to $\ell'$ can be seen as proposing to flip the lower order bit in a binary representation of $\ell'$. For example, a proposal to move from $\ell = 3$ to $\ell = 2$ can be seen as proposing to change $\ell$ from $11$ to $10$ (in binary). Such a proposal will always be accepted assuming a transition density for which $p(x_{i}|x_{i-1}) = p(\textnormal{--}x_{i}|\textnormal{--}x_{i-1})$ and an observation probability which depends on $x_{i}$ only via $|x_{i}|$.

\subsection{Relation to PGBS}

The forward pool state selection scheme can be used to construct a sampler with properties similar to PGBS. This is done by using independence Metropolis to sample values of $x$ and $\ell$ from $\lambda_{i}$. 

At time $1$, we propose our pool states from $p(x)$. At times $i > 2$, we propose $\ell'$ by selecting it uniformly at random from $\{1, \ldots, L\}$ and we propose $x'$ by sampling from $p(x|x_{i-1}^{[\ell']})$. The proposals at all times $i$ are accepted with probability
\begin{eqnarray}
\min\biggl(1, \frac{p(y_{i}|x_{i}')}{p(y_{i}|x_{i})}\biggr)
\end{eqnarray}
This sampler has computational cost proportional to $Ln$ per iteration, like PGBS. It is analogous to a PGBS sampler with importance densities
\begin{eqnarray}
q_{1}(x|y_{1}) = p(x)
\end{eqnarray}
and
\begin{eqnarray}
q_{i}(x|x_{i-1}, y_{i}) = p(x|x_{i-1}), \quad i > 2
\end{eqnarray}
with the key difference between these two samplers being that PGBS uses importance weights $p(y_{i}|x_{i})$ on each particle, instead of an independence Metropolis accept-reject step.

\section{Proof of correctness}

We modify the original proof of Neal (2003), which assumes that the sets of pool states $\mathcal{P}_{1}, \ldots, \mathcal{P}_{n}$ are selected independently at each time, to show the validity of our new sequential pool state selection scheme. Another change in the proof is to account for generating the pool states by sampling them from $\lambda_{i}$ or $\gamma_{i}$ instead of $\kappa_{i}^{f}$ or $\kappa_{i}^{b}$.

This proof shows that the probability of starting at $x$ and moving to $x'$ with given sets of pool states $\mathcal{P}_{i}$ (consisting of values of $x$ at each time $i$), pool indices $l_{i}$ of $x_{i}$, and pool indices $l_{i}'$ of $x_{i}'$ is the same as the probability of starting at $x'$ and moving to $x$ with the same set of pool states $\mathcal{P}_{i}$, pool indices $l_{i}'$ of $x_{i}'$, and pool indices $l_{i}$ of $x_{i}$. This in turn implies, by summing/integrating over $\mathcal{P}_{i}$ and $l_{i}$, that the embedded HMM method with the sequential pool state scheme satisfies detailed balance with respect to $p(x|y)$, and hence leaves $p(x|y)$ invariant.

Suppose we use the sequential forward scheme. The probability of starting at $x$ and moving to $x'$ decomposes into the product of the probability of starting at $x$, which is $p(x|y)$, the probability of choosing a set of pool state indices $l_{i}$, which is $\frac{1}{L^{n}}$, the probability of selecting the initial values of $\ell_{i}$ for the stochastic initialization step, the probability of selecting the sets of pool states $\mathcal{P}_{i}$, $P(\mathcal{P}_{1}, \ldots, \mathcal{P}_{n})$, and finally the probability of choosing $x'$.

The probability of selecting given initial values for the links to previous states $\ell_{2}, \ldots, \ell_{n}$ is
\begin{eqnarray}
\prod_{i=2}^{n}\frac{p(x_{i}|x_{i-1}^{[\ell_{i}]})}{\sum_{m=1}^{L}p(x_{i}|x_{i-1}^{[m]})}
\end{eqnarray}
The probability of choosing a given set of pool states is
\begin{eqnarray}
P(\mathcal{P}_{1}, \ldots, \mathcal{P}_{n}) = P(\mathcal{P}_{1})\prod_{i=2}^{n}P(\mathcal{P}_{i}|\mathcal{P}_{i-1})
\end{eqnarray}
At time $1$, we use a Markov chain with invariant density $\kappa_{1}$ to select pool states in $\mathcal{P}_{1}$. Therefore
\begin{eqnarray}
P(\mathcal{P}_{1}) &=& \prod_{j=l_{1}+1}^{L}R_{1}(x_{1}^{[j]}|x_{1}^{[j-1]})\prod_{j=l_{1}-1}^{1}\tilde{R}_{1}(x_{1}^{[j]}|x_{1}^{[j+1]}) \notag \\
&=& \prod_{j=l_{1}+1}^{L}R_{1}(x_{1}^{[j]}|x_{1}^{[j-1]})\prod_{j=l_{1}-1}^{1}R_{1}(x_{1}^{[j+1]}|x_{1}^{[j]})\frac{\kappa_{1}(x_{1}^{[j]})}{\kappa_{1}(x_{1}^{[j+1]})} \notag \\
&=&\prod_{j=l_{1}}^{L-1}R_{1}(x_{1}^{[j+1]}|x_{1}^{[j]})\prod_{j=l_{1}-1}^{1}R_{1}(x_{1}^{[j+1]}|x_{1}^{[j]})\frac{\kappa_{1}(x_{1}^{[j]})}{\kappa_{1}(x_{1}^{[j+1]})} \notag \\
&=&\frac{\kappa_{1}(x_{1}^{[1]})}{\kappa_{1}(x_{1}^{[l_{1}]})}\prod_{j=1}^{L-1}R_{1}(x_{1}^{[j+1]}|x_{1}^{[j]})
\end{eqnarray}
For times $i > 1$ we use a Markov chain with invariant density $\lambda_{i}$ to sample a set of pool states, given $\mathcal{P}_{i-1}$. The chain is started at $x_{i}^{[l_{i}]} = x_{i}$ and $\ell_{i}^{[l_{i}]} = \ell_{i}$. Therefore
\begin{eqnarray}
P(\mathcal{P}_{i} | \mathcal{P}_{i-1}) &=& \prod_{j=l_{1}+1}^{L}R_{i}(x_{i}^{[j]}, \ell_{i}^{[j]}|x_{i}^{[j-1]}, \ell_{i}^{[j-1]})\prod_{j=l_{i}-1}^{1}\tilde{R}_{i}(x_{i}^{[j]}, \ell_{i}^{[j]}|x_{i}^{[j+1]}, \ell_{i}^{[j+1]}) \notag \\
&=& \prod_{j=l_{i}+1}^{L}R_{i}(x_{i}^{[j]}, \ell_{i}^{[j]}|x_{i}^{[j-1]}, \ell_{i}^{[j-1]})\prod_{j=l_{i}-1}^{1}R_{i}(x_{1}^{[j+1]}, \ell_{i}^{[j+1]}|x_{i}^{[j]}, \ell_{i}^{[j]})\frac{\lambda_{i}(x_{i}^{[j]}, \ell_{i}^{[j]})}{\lambda_{i}(x_{1}^{[j+1]}, \ell_{i}^{[j+1]})} \notag \\
&=&\prod_{j=l_{i}}^{L-1}R_{i}(x_{i}^{[j+1]}, \ell_{i}^{[j+1]}|x_{i}^{[j]}, \ell_{i}^{[j]})\prod_{j=l_{i}-1}^{1}R_{i}(x_{i}^{[j+1]}, \ell_{i}^{[j+1]}|x_{i}^{[j]}, \ell_{i}^{[j]})\frac{\lambda_{i}(x_{i}^{[j]}, \ell_{i}^{[j]})}{\lambda_{i}(x_{i}^{[j+1]}, \ell_{i}^{[j+1]})} \notag \\
&=& \frac{\lambda_{i}(x_{i}^{[1]}, \ell_{i}^{[1]})}{\lambda_{i}(x_{i}^{[l_{i}]}, \ell_{i}^{[l_{i}]})}\prod_{j=1}^{L-1}R_{i}(x_{i}^{[j+1]}, \ell_{i}^{[j+1]}|x_{i}^{[j]}, \ell_{i}^{[j]})
\end{eqnarray}

Finally, we choose a new sequence $x'$ amongst the collection of sequences consisting of the pool states with a backward pass. This is done by first choosing a pool state $x_{n}'$ uniformly at random from $\mathcal{P}_{n}$. We then select the remaining states $x_{i}^{[l_{i}']}$ by selecting $l_{1}', \ldots, l_{n-1}'$ with probability
\begin{eqnarray}
\prod_{i=2}^{n}\frac{p(x_{i}'|x_{i-1}^{[l_{i-1}']})}{\sum_{m=1}^{L}p(x_{i}'|x_{i-1}^{[m]})}
\end{eqnarray}

Thus, the probability of starting at $x$ and going to $x'$, with given $\mathcal{P}_{1}, \ldots, \mathcal{P}_{n}$, $l_{1}, \ldots, l_{n}$ and $l_{1}', \ldots, l_{n}'$ is
\begin{eqnarray}
&&p(x|y) \times \frac{1}{L^{n}} \times \prod_{i=2}^{n}\frac{p(x_{i}|x_{i-1}^{[\ell_{i}]})}{\sum_{m=1}^{L}p(x_{i}|x_{i-1}^{[m]})} \times \frac{\kappa_{1}(x_{1}^{[1]})}{\kappa_{1}(x_{1}^{[l_{1}]})}\prod_{j=1}^{L-1}R_{1}(x_{1}^{[j+1]}|x_{1}^{[j]}) 
\label{eq:proba} \\
&& \ \times \ \prod_{i=2}^{n}\biggl[\frac{\lambda_{i}(x_{i}^{[1]}, \ell_{i}^{[1]})}{\lambda_{i}(x_{i}^{[l_{i}]}, \ell_{i}^{[l_{i}]})}\prod_{j=1}^{L-1}R_{i}(x_{i}^{[j+1]}, \ell_{i}^{[j+1]}|x_{i}^{[j]}, \ell_{i}^{[j]})\biggr]\times \frac{1}{L} \times \prod_{i=2}^{n}\frac{p(x_{i}'|x_{i-1}^{[l_{i-1}']})}{\sum_{m=1}^{L}p(x_{i}'|x_{i-1}^{[m]})} \notag \\
&&= \ \kappa_{1}(x_{1}^{[1]})\prod_{j=1}^{L-1}R_{1}(x_{1}^{[j+1]}|x_{1}^{[j]}) \times \ \prod_{i=2}^{n}\biggl[\lambda_{i}(x_{i}^{[1]}, \ell_{i}^{[1]})\prod_{j=1}^{L-1}R_{i}(x_{i}^{[j+1]}, \ell_{i}^{[j+1]}|x_{i}^{[j]}, \ell_{i}^{[j]})\biggr]\notag \\
&& \ \times \ \frac{1}{L^{n+1}} \times \frac{p(x|y)}{\kappa_{1}(x_{1})\prod_{i=2}^{n}\lambda_{i}(x_{i}, \ell_{i})} \times \prod_{i=2}^{n}\frac{p(x_{i}|x_{i-1}^{[\ell_{i}]})}{\sum_{m=1}^{L}p(x_{i}|x_{i-1}^{[m]})}\prod_{i=2}^{n}\frac{p(x_{i}'|x_{i-1}^{[l_{i-1}']})}{\sum_{m=1}^{L}p(x_{i}'|x_{i-1}^{[m]})} \notag
\end{eqnarray}
Here, we have $x_{i-1}^{[l_{i-1}']} = x_{i-1}'$. Also $\kappa_{1}(x_{1}) = p(x_{1})p(y_{1}|x_{1})/\sum_{x_{1} \in \mathcal{P}_{1}}p(x_{1})p(y_{1}|x_{1})$ and
\begin{eqnarray}
\prod_{i=2}^{n}\lambda_{i}(x_{i}, \ell_{i}) = \prod_{i=2}^{n}\frac{p(y_{i}|x_{i})p(x_{i}|x_{i-1}^{[\ell_{i}]})}{\sum_{x_{i} \in \mathcal{P}_{i}} \sum_{m=1}^{L}p(y_{i}|x_{i})p(x_{i}|x_{i-1}^{[m]})}
\end{eqnarray}
and
\begin{eqnarray}
p(x|y) = \frac{p(x_{1})\prod_{i=2}^{n}p(x_{i}|x_{i-1})\prod_{i=1}^{n}p(y_{i}|x_{i})}{p(y)}
\end{eqnarray}
Therefore (\ref{eq:proba}) can be simplified to
\begin{eqnarray}
&& \frac{1}{p(y)}\kappa_{1}(x_{1}^{[1]})\prod_{j=1}^{L-1}R_{1}(x_{1}^{[j+1]}|x_{1}^{[j]}) \times \ \prod_{i=2}^{n}\biggl[\lambda_{i}(x_{i}^{[1]}, \ell_{i}^{[1]})\prod_{j=1}^{L-1}R_{i}(x_{i}^{[j+1]}, \ell_{i}^{[j+1]}|x_{i}^{[j]}, \ell_{i}^{[j]})\biggr] \times \frac{1}{L^{n+1}} \notag \\
&& \times \ \prod_{i=2}^{n}p(x_{i}|x_{i-1}) \times \prod_{i=2}^{n}p(x_{i}'|x_{i-1}') \times \prod_{i=2}^{n}\frac{1}{\sum_{m=1}^{L}p(x_{i}|x_{i-1}^{[m]})} \times \prod_{i=2}^{n}\frac{1}{\sum_{m=1}^{L}p(x_{i}'|x_{i-1}^{[m]})} \notag \\
&& \times \ \sum_{x_{1} \in \mathcal{P}_{1}}p(x_{1})p(y_{1}|x_{1})\prod_{i=2}^{n}\sum_{x_{i} \in \mathcal{P}_{i}} \sum_{m=1}^{L}p(y_{i}|x_{i})p(x_{i}|x_{i-1}^{[m]})
\end{eqnarray}
The last factor in the product only depends on the selected set of pool states. By exchanging $x$ and $x'$ we see that the probability of starting at $x'$ and then going to $x$, with given sets of pool states $\mathcal{P}_{i}$, pool indices $l_{i}$ of $x_{i}$ and pool indices $l_{i}'$ of $x_{i}'$ is the same.

\section{Experiments}

\subsection{Test models}

To demonstrate the performance of our new pool state scheme, we use two different state space models. The latent process for both models is a vector autoregressive process, with
\begin{eqnarray}
X_{1} &\sim& N(0, \Sigma_{\textnormal{init}}) \\
X_{i} | x_{i-1} &\sim& N(\Phi x_{i-1}, \Sigma), \quad i = 2, \ldots, n
\end{eqnarray}
where $X_{i} = (X_{i, 1}, \ldots, X_{i, P})'$ and
\begin{eqnarray}
\Phi &=& 
\begin{pmatrix}
\phi_{1} & \ldots & 0 \\
\vdots & \ddots & \vdots \\
0 & \ldots & \phi_{P} \\
\end{pmatrix} \\
\Sigma &=& 
\begin{pmatrix}
1 & \ldots & \rho \\
\vdots & \ddots & \vdots \\
\rho & \ldots & 1 \\
\end{pmatrix} \\
\Sigma_{\textnormal{init}} &=& 
\begin{pmatrix}
\frac{1}{1 - \phi_{1}^{2}} & \ldots & \frac{\rho}{\sqrt{1 - \phi_{1}^{2}}\sqrt{1 - \phi_{P}^{2}}} \\
\vdots & \ddots & \vdots \\
\frac{\rho}{\sqrt{1 - \phi_{P}^{2}}\sqrt{1 - \phi_{1}^{2}}} & \ldots & \frac{1}{1 - \phi_{P}^{2}} \\
\end{pmatrix}
\end{eqnarray}
Note that $\Sigma_{\textnormal{init}}$ is the covariance of the stationary distribution for this process.

For model 1, the observations are given by
\begin{eqnarray}
Y_{i, j} | x_{i, j} &\sim & \mbox{Poisson}(\exp(c_{j} + \sigma_{j} x_{i, j})), \quad i = 1, \ldots, n, \quad j = 1, \ldots, P
\end{eqnarray}

For model 2, the observations are given by
\begin{eqnarray}
Y_{i, j} | x_{i, j} &\sim & \mbox{Poisson}(\sigma_{j} |x_{i, j}|), \quad i = 1, \ldots, n, \quad j = 1, \ldots, P
\end{eqnarray}

For model 1, we use a $10$-dimensional latent state and a sequence length of $n = 250$, setting parameter values to $\rho = 0.7$, and $c_{j} = -0.4$, $\phi_{j} = 0.9$, $\sigma_{j} = 0.6$ for $j = 1, \ldots, P$, with $P = 10$.

For model 2, we increase the dimensionality of the latent space to $15$ and the sequence length to $500$. We set $\rho = 0.7$ and $\phi_{j} = 0.9$, $\sigma_{j} = 0.8$ for $j = 1, \ldots, P$, with $P = 15$.  

\begin{figure}[t]
         \centering
         \begin{subfigure}[b]{0.45\textwidth}
                 \includegraphics[width=\textwidth]{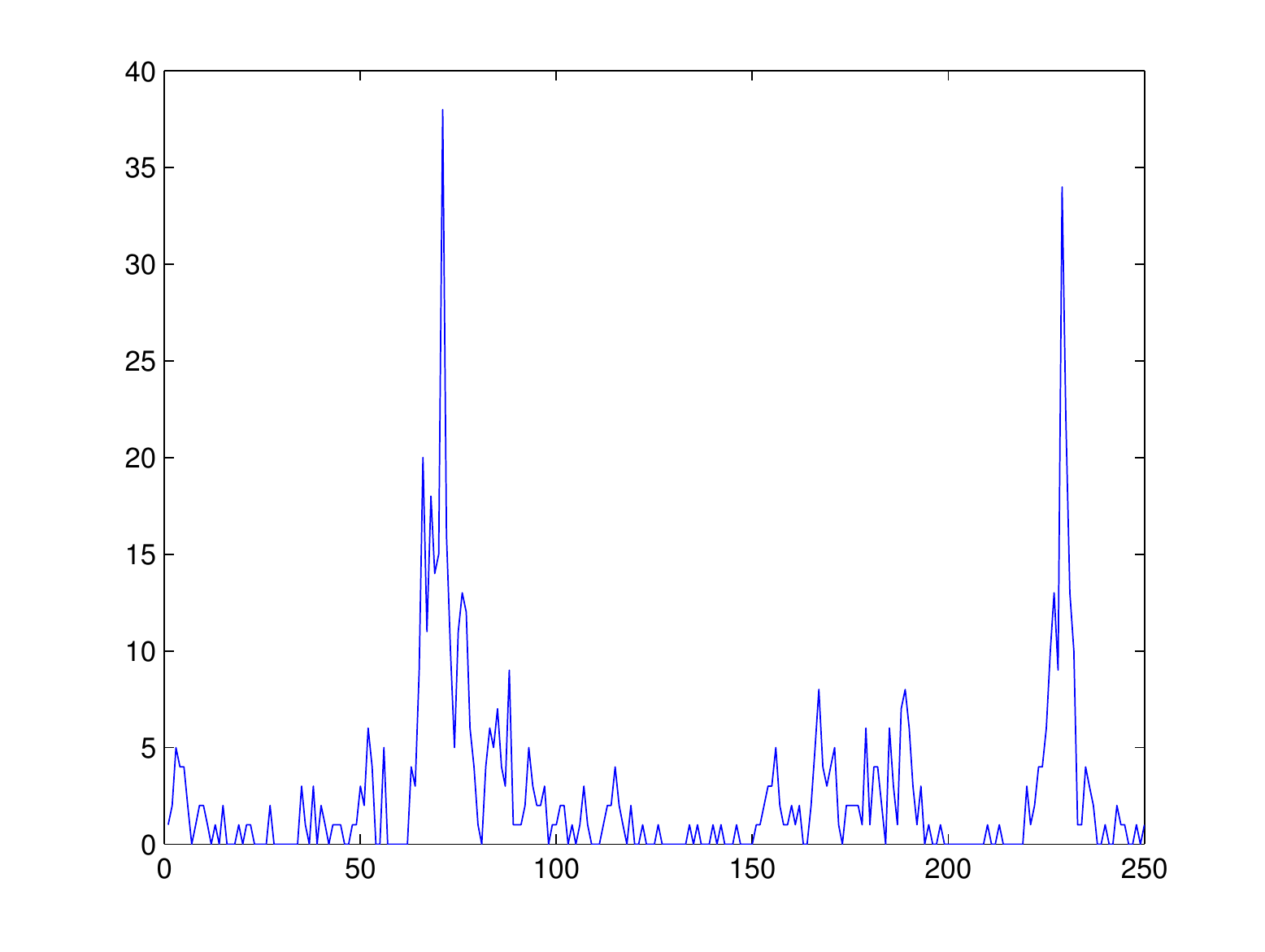}
                 \caption{Model 1}
         \end{subfigure}
	 ~
         \begin{subfigure}[b]{0.45\textwidth}
                 \includegraphics[width=\textwidth]{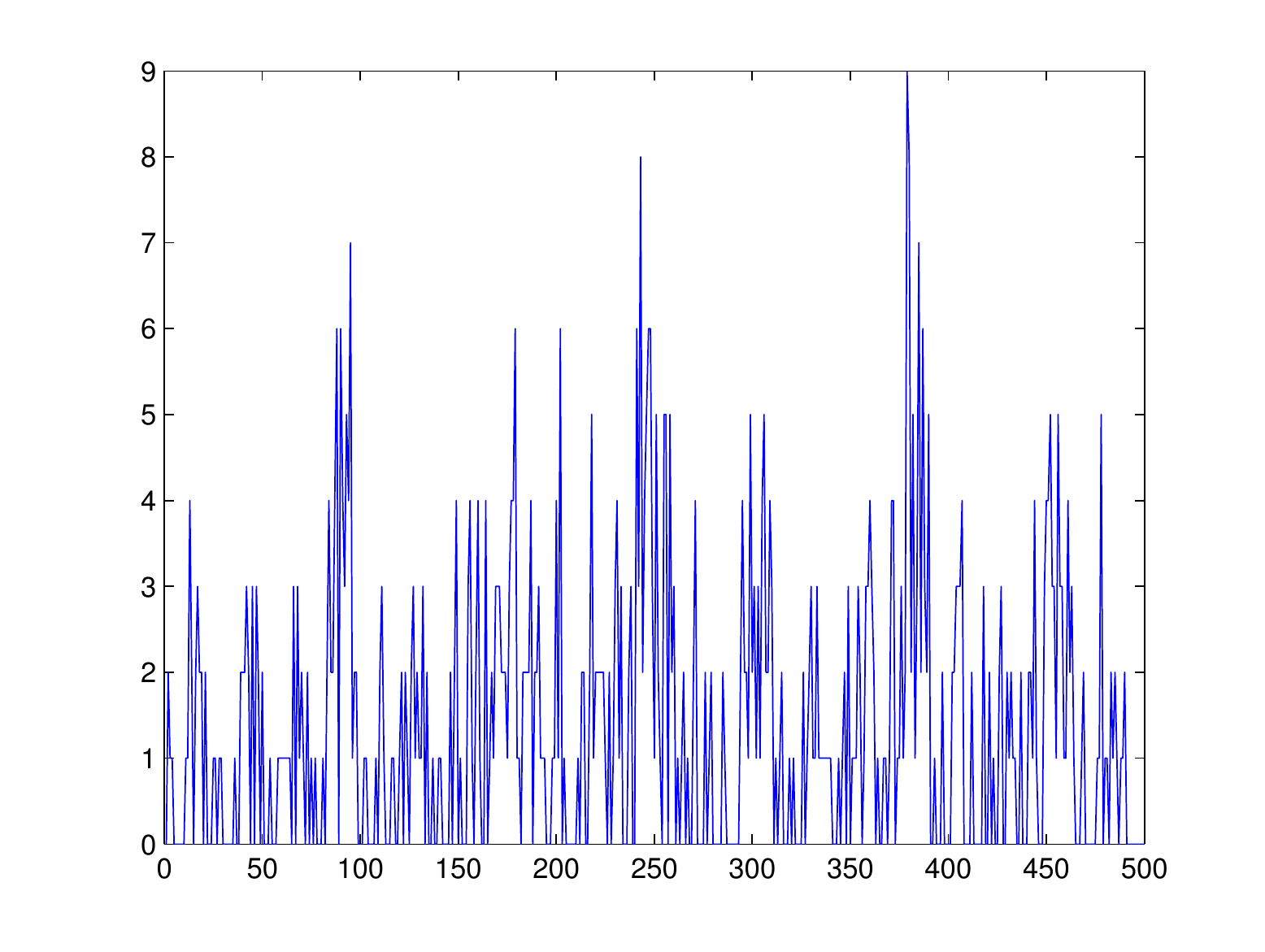}
                 \caption{Model 2}
         \end{subfigure}
         \caption{Observations from Model 1 and Model 2 along dimension $j = 1$.}
         \label{fig:data}
\end{figure}

We generated one random sequence from each model to test our samplers on. These observations from model 1 and model 2 are shown in Figure \ref{fig:data}. Note that we are testing only sampling of the latent  variables, with the parameters set to their true values. The code for all experiments in this paper is available at http://arxiv.org/abs/1602.06030.

\subsection{Single-state Metropolis Sampler}

A simple scheme for sampling the latent state sequence is to use Metropolis-Hastings updates that sample each $x_{i}$ in sequence, conditional on $x_{-i} = (x_{1}, \ldots, x_{i-1}, x_{i+1}, \ldots, x_{i})$ and the data, starting at time $1$ and going to time $n$. We sample all dimensions of $x_{i}$ at once using autoregressive updates (see section 5.4.3).

The conditional densities of the $X_{i}$ are
\begin{eqnarray*}
p(x_{1}|x_{-1}, y) &\propto & \ p(x_{1} | x_{2})p(y_{1}|x_{1}) \ \propto \ p(x_{1})p(x_{2}|x_{1})p(y_{1}|x_{1}) \\
p(x_{i}|x_{-i}, y) &\propto & p(x_{i} | x_{i-1}, x_{i+1})p(y_{i}|x_{i}) \ \propto \ p(x_{i}|x_{i-1})p(x_{i+1}|x_{i})p(y_{i}|x_{i}), \quad 2 \leq i \leq n-1 \\
p(x_{n}|x_{-n}, y) &\propto & p(x_{n}|x_{n-1})p(y_{n}|x_{n})
\end{eqnarray*}

The densities $p(x_{1} | x_{2}), p(x_{i} | x_{i-1}, x_{i+1})$, and $p(x_{n}|x_{n-1})$ are all Gaussian. The means and covariances for these densities can be derived by viewing $p(x_{1})$ or $p(x_{i}|x_{i-1})$ as a Gaussian prior for $x_{i}$ and $p(x_{i+1}|x_{i})$ as a Gaussian likelihood for $x_{i}$. In particular, we have
\begin{eqnarray*}
X_{1} | x_{2} &\sim & N(\mu_{1}, \Sigma_{1}) \\
X_{i} | x_{i}, x_{i+1} &\sim & N(\mu_{i}, \Sigma_{i}) \\
X_{n} | x_{n-1} &\sim & N(\mu_{n}, \Sigma_{n})
\end{eqnarray*}
where
\begin{eqnarray*}
\mu_{1} &=& [(\Phi^{-1}\Sigma\Phi^{-1})^{-1} + \Sigma_{\textnormal{init}}^{-1}]^{-1}[(\Phi^{-1}\Sigma\Phi^{-1})^{-1}\Phi^{-1}x_{2}] \\
&=& [(\Phi^{-1}\Sigma\Phi^{-1})^{-1} + \Sigma_{\textnormal{init}}^{-1}]^{-1}[\Sigma^{-1}(\Phi x_{2})] \\
&=& [\Phi^{2} + \Sigma_{\textnormal{init}}^{-1}\Sigma]^{-1}\Phi x_{2} \\
\Sigma_{1} &=& [(\Phi^{-1}\Sigma\Phi^{-1})^{-1} + \Sigma_{\textnormal{init}}^{-1}]^{-1}\\ 
&=& [\Phi(\Sigma^{-1}\Phi) + \Sigma_{\textnormal{init}}^{-1}]^{-1} \\ \\ 
\mu_{i} &=& [(\Phi^{-1}\Sigma\Phi^{-1})^{-1} + \Sigma^{-1}]^{-1}[\Sigma^{-1}\Phi x_{i-1} + (\Phi^{-1}\Sigma\Phi^{-1})^{-1}\Phi^{-1}x_{i+1}] \\
&=& (\Phi^{-1}\Sigma\Phi^{-1})^{-1} + \Sigma^{-1}]^{-1}[\Sigma^{-1}(\Phi (x_{i-1} + x_{i+1}))] \\
&=& [\Phi^{2} + I]^{-1}\Phi (x_{i-1} + x_{i+1}) \\
\Sigma_{i} &=& [(\Phi^{-1}\Sigma\Phi^{-1})^{-1} + \Sigma^{-1}]^{-1} \\
&=& [\Phi(\Sigma^{-1}\Phi) + \Sigma^{-1}]^{-1} \\ \\
\mu_{n} &=& \Phi x_{n-1} \\
\Sigma_{n} &=& \Sigma
\end{eqnarray*}
To speed up the Metropolis updates, we precompute and store the matrices $[\Phi^{2} + \Sigma_{\textnormal{init}}^{-1}\Sigma]^{-1}\Phi$, $[\Phi^{2} + I]^{-1}\Phi$ as well as the Cholesky decompositions of the posterior covariances.

In both of our test models, the posterior standard deviation of the latent variables $x_{i,j}$ varies depending on the value of the observed $y_{i,j}$. To address this, we alternately use a larger or a smaller proposal scaling, $\epsilon$, in the autoregressive update when performing an iteration of the Metropolis sampler.

\subsection{Particle Gibbs with Backward Sampling with Metropolis}

We implement the PGBS method as described in Section 5.3, using the initial density $p(x)$ and the transition densities $p(x_{i}|x_{i-1})$ as importance densities to generate particles. We combine PGBS updates with single-state Metropolis updates from Section 5.6.2. This way, we combine the strengths of the two samplers in targeting different parts of the posterior distribution. In particular, we expect the Metropolis updates to do better for the $x_{i}$ with highly informative $y_{i}$, and the PGBS updates to do better for the $x_{i}$ where $y_{i}$ is not as informative.

\subsection{Tuning the Baseline Samplers}

For model $1$, we compared the embedded HMM sampler to the simple single-state Metropolis sampler, to the PGBS sampler, and to the combination of PGBS with Metropolis. For model 2, we compared the embedded HMM sampler to the PGBS with Metropolis sampler. For both models and all samplers, we ran the sampler five times using five different random number generator seeds. We implemented the samplers in MATLAB on a Linux system with a 2.60 GHz Intel i7-3720QM CPU.

\subsubsection{Model 1}

For the single-state Metropolis sampler, we initialized all $x_{i,j}$ to $0$. Every iteration alternately used a scaling factor, $\epsilon$, of either $0.2$ or $0.8$, which resulted in an average acceptance rate of between $30\%$ and $90\%$ for the different $x_{i}$ over the sampler run. We ran the sampler for $1000000$ iterations, and prior to analysis, the resulting sample was thinned by a factor of $10$, to $100 000$. The thinning was done due to the difficulty of working with all samples at once, and after thinning the samples still possessed autocorrelation times significantly greater than $1$. Each of the $100000$ samples took about $0.17$ seconds to draw.

For the PGBS sampler and the sampler combining PGBS and Metropolis updates, we also initialized all $x_{i,j}$ to $0$. We used $250$ particles for the PGBS updates. For the Metropolis updates, we alternated between scaling factors of $0.2$ and $0.8$, which also gave acceptance rates between $30\%$ and $90\%$. For the standalone PGBS sampler, we performed a total of $70000$ iterations. Each iteration produced two samples for a total of $140000$ samples and consisted of a PGBS update using the forward sequence and a PGBS update using the reversed sequence. Each sample took about $0.12$ seconds to draw. For the PGBS with Metropolis sampler, we performed a total of $30000$ iterations of the sampler. Each iteration was used to produce four samples, for a total of $120000$ samples, and consisted of a PGBS update using the forward sequence, ten Metropolis updates (of which only the value after the tenth update was retained), a PGBS update using the reversed sequence, and another ten Metropolis updates, again only keeping the value after the tenth update. The average time to draw each of the $120000$ samples was about $0.14$ seconds.

\subsubsection{Model 2}

For model 2, we were unable to make the single-state Metropolis sampler converge to anything resembling the actual posterior in a reasonable amount of time. In particular, we found that for $x_{i,j}$ sufficiently far from $0$, the Metropolis sampler tended to be stuck in a single mode, never visiting values with the opposite sign.

For the PGBS with Metropolis sampler, we set the initial values of $x_{i, j}$ to $1$. We set the number of particles for PGBS to $80000$, which was nearly the maximum possible for the memory capacity of the computer we used. For the Metropolis sampler, we alternated between scaling factors of $0.3$ and $1$, which resulted in acceptance rates ranging between $29\%$ and $72\%$. We performed a total of $250$ iterations of the sampler. As for model $1$, each iteration produced four samples, for a total of $1000$ samples, and consisted of a PGBS update with the forward sequence, fifty Metropolis updates (of which we only keep the value after the last one), a PGBS update using the reversed sequence, and another fifty Metropolis updates (again only keeping the last value). It took about $26$ seconds to draw each sample.

\subsection{Embedded HMM sampling}

For both model 1 and model 2, we implemented the proposed embedded HMM method using the forward pool state selection scheme, alternating between updates that use the original and the reversed sequence. As for the baseline samplers, we ran the embedded HMM samplers five times for both models, using five different random number generator seeds. 

We generate pool states at time $1$ using autoregressive updates to sample from $\kappa_{1}^{f}$. At times $i \geq 2$, we sample each pool state from $\lambda_{i}(x, l)$ by combining an autoregressive and shift update. The autoregressive update proposes to only change $x$, keeping the current $l$ fixed. The shift update samples both $x$ and $l$, with a new $l$ proposed from a Uniform$\{1, \ldots, L\}$ distribution. For model 2, we also add a flip update to generate a negated version of each pool state.

Note that the chain used to produce the pool states now uses a sequence of updates. Therefore, if our forward transition first does an autoregressive update and then a shift update, the reverse transitions must first do a shift update and then an autoregressive update.

As for the single-state Metropolis updates, it is beneficial to use a different proposal scaling, $\epsilon$, when generating each pool state at each time $i$. This allows generation of sets of pool states which are more concentrated when $y_{i}$ is informative and more dispersed when $y_{i}$ holds little information.

\subsubsection{Model 1}

For model 1, we initialized all $x_{i,j}$ to $0$. We used $50$ pool states for the embedded HMM updates. For each Metropolis update to sample a pool state, we used a different scaling $\epsilon$, chosen at random from a $\textnormal{Uniform}(0.1, 0.4)$ distribution. The acceptance rates ranged between $55\%$ and $95\%$ for the Metropolis updates and between $20\%$ and $70\%$ for the shift updates. We performed a total of $9000$ iterations of the sampler, with each iteration consisting of an embedded HMM update using the forward sequence and an embedded HMM update using the reversed sequence, for a total of $18000$ samples. Each sample took about $0.81$ seconds to draw.

\subsubsection{Model 2}

For model 2, we initialized the $x_{i,j}$ to $1$. We used a total of $80$ pool states for the embedded HMM sampler (i.e.\ $40$ positive-negative pairs due to flip updates). Each Metropolis update used to sample a pool state used a scaling, $\epsilon$, randomly drawn from the $\textnormal{Uniform}(0.05, 0.2)$ distribution. The acceptance rates ranged between $75\%$ and $90\%$ for the Metropolis updates and between $20\%$ and $40\%$ for the shift updates. We performed a total of $9000$ iterations of the sampler, producing two samples per iteration with an embedded HMM update using the forward sequence and an embedded HMM update using the reversed sequence. Each of the $18000$ samples took about $1.4$ seconds to draw.

\subsection{Comparisons}

As a way of comparing the performance of the two methods, we use an estimate of autocorrelation time\footnote{Technically, when we alternate updates with the forward and reversed sequence or mix PGBS and single-state Metropolis updates, we cannot use autocorrelation times to measure how well the chain explores the space. While the sampling scheme leaves the correct target distribution invariant, the flipping of the sequence makes the sampling chain for a given variable non-homogeneous. However, suppose that instead of deterministically flipping the sequence at every step, we add an auxiliary indicator variable that determines (given the current state) whether the forward or the reversed sequence is used, and that the probability of flipping this indicator variable is nearly one. With this auxiliary variable the sampling chain becomes homogeneous, with its behaviour nearly identical to that of our proposed scheme. Using autocorrelation time estimates to evaluate the performance of our sampler is therefore valid, for all practical purposes.} for each of the latent variables $x_{i, j}$. Autocorrelation time is a measure of how many draws need to be made using the sampling chain to produce the equivalent of one independent sample. The autocorrelation time is defined as $\tau = 1 + 2\sum_{i=1}^{\infty} \rho_{k}$, where $\rho_{k}$ is the autocorrelation at lag $k$. It is commonly estimated as
\begin{eqnarray}
\hat{\tau} = 1 + 2\sum_{i=1}^{K}\hat{\rho}_{k}
\end{eqnarray}
where $\hat{\rho}_{k}$ are estimates of lag-$k$ autocorrelations and the cutoff point $K$ is chosen so that $\hat{\rho}_{k}$ is negligibly different from $0$ for $k > K$. Here
\begin{eqnarray}
\hat{\rho}_{k} = \frac{\hat{\gamma}_{k}}{\hat{\gamma}_{0}}
\end{eqnarray}
where $\hat{\gamma}_{k}$ is an estimate of the lag-$k$ autocovariance
\begin{eqnarray}
\hat{\gamma}_{k} = \frac{1}{n}\sum_{l = 1}^{n-k}(x_{l} - \bar{x})(x_{k+l} - \bar{x})
\end{eqnarray}
When estimating autocorrelation time, we remove the first $10\%$ of the sample as burn-in. Then, to estimate $\hat{\gamma}_{k}$, we first estimate autocovariances for each of the five runs, taking $\bar{x}$ to be the overall mean over the five runs. We then average these five autocovariance estimates to produce $\hat{\gamma}_{k}$. To speed up autocovariance computations, we use the Fast Fourier Transform. The autocorrelation estimates are then adjusted for computation time, by multiplying the estimated autocorrelation time by the time it takes to draw a sample, to ensure that the samplers are compared fairly.

\begin{figure}[t]
         \centering
         \begin{subfigure}[b]{0.45\textwidth}
                 \includegraphics[width=\textwidth]{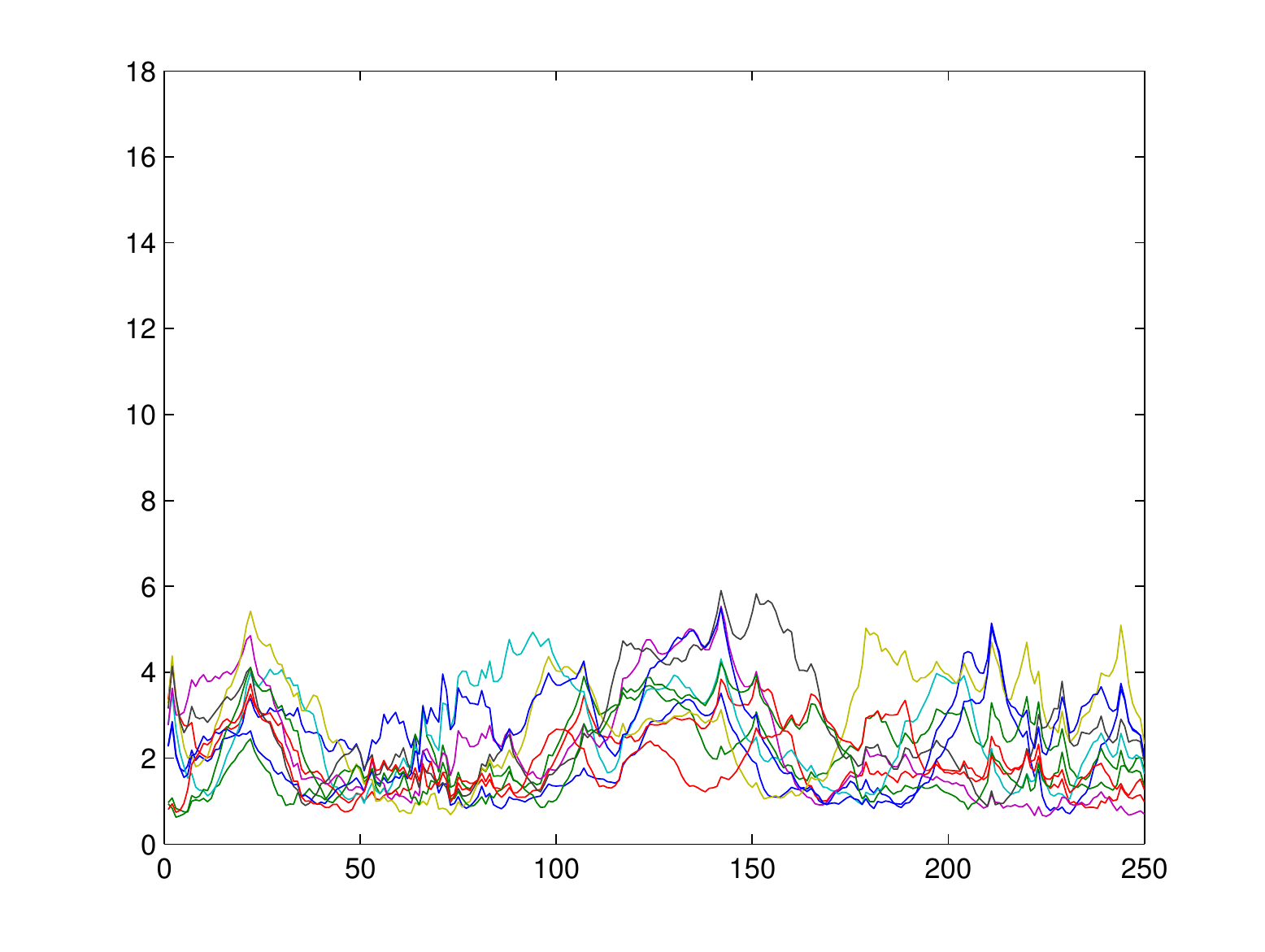}
                 \caption{Metropolis (0.17 seconds/sample)}
         \end{subfigure}
	 ~
         \begin{subfigure}[b]{0.45\textwidth}
                 \includegraphics[width=\textwidth]{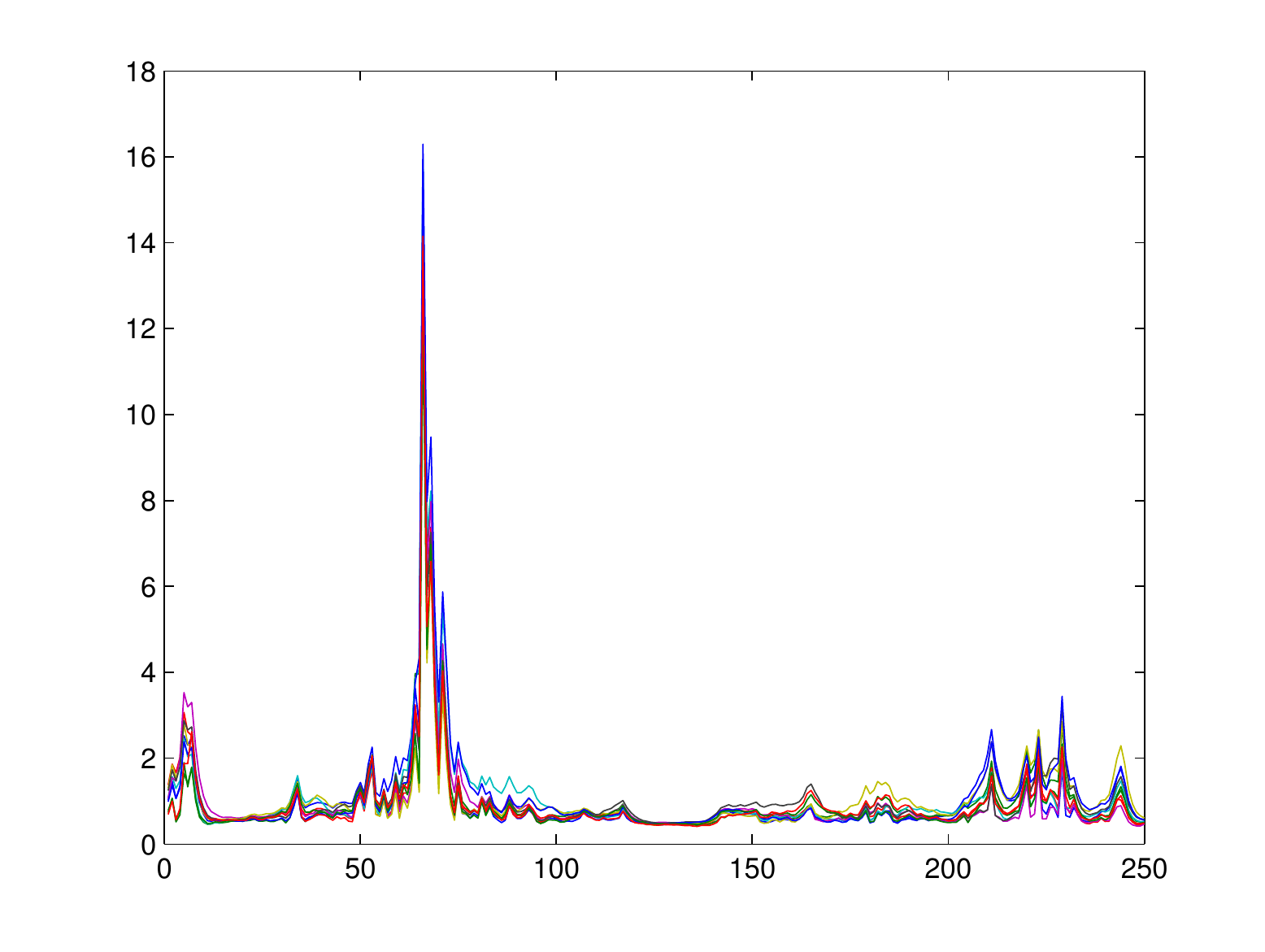}
                 \caption{PGBS (0.12 seconds/sample)}
         \end{subfigure}
         \begin{subfigure}[b]{0.45\textwidth}
                 \includegraphics[width=\textwidth]{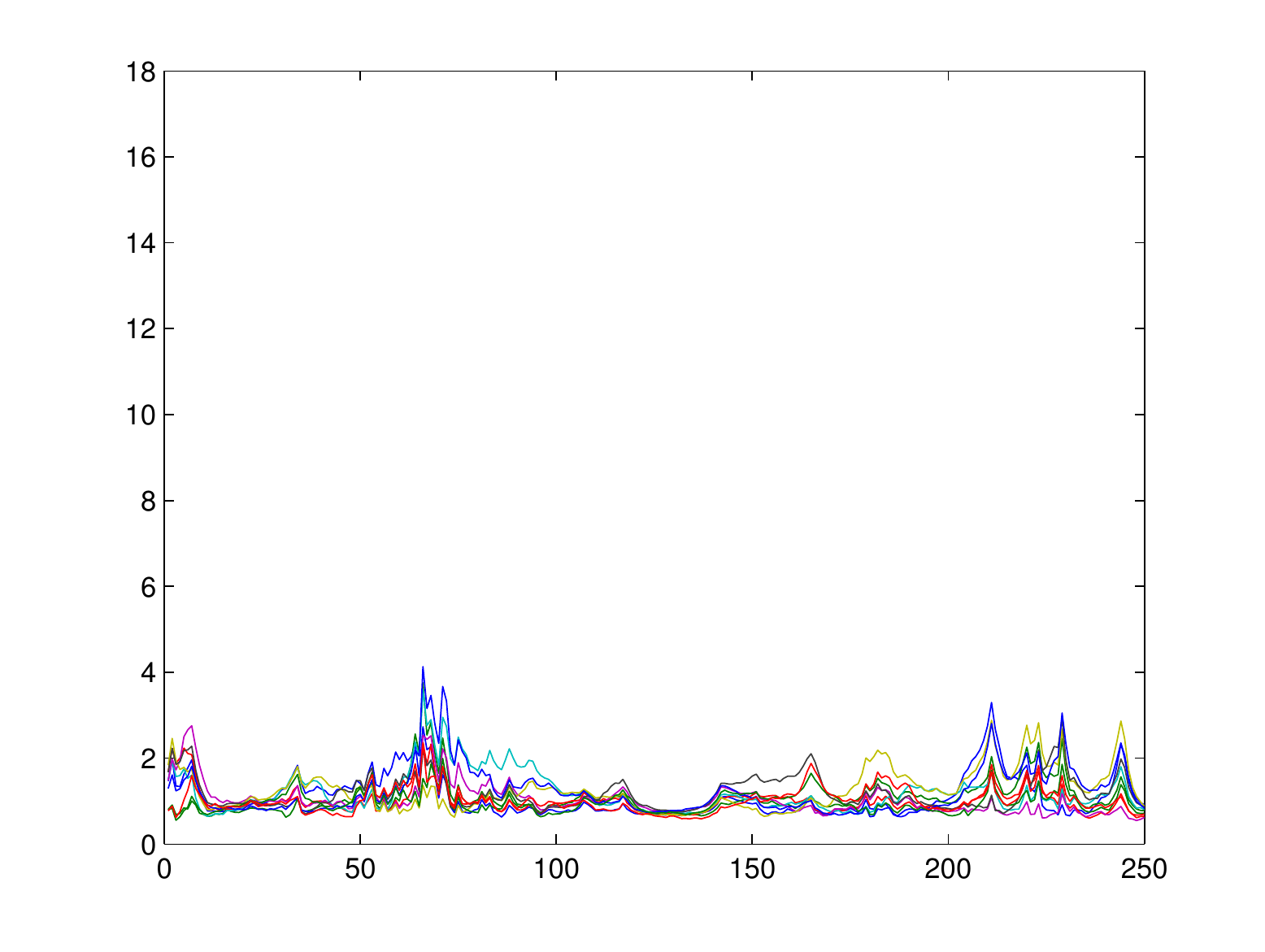}
                 \caption{PGBS+Metropolis (0.14 seconds/sample)}
         \end{subfigure}
     ~
         \begin{subfigure}[b]{0.45\textwidth}
                 \includegraphics[width=\textwidth]{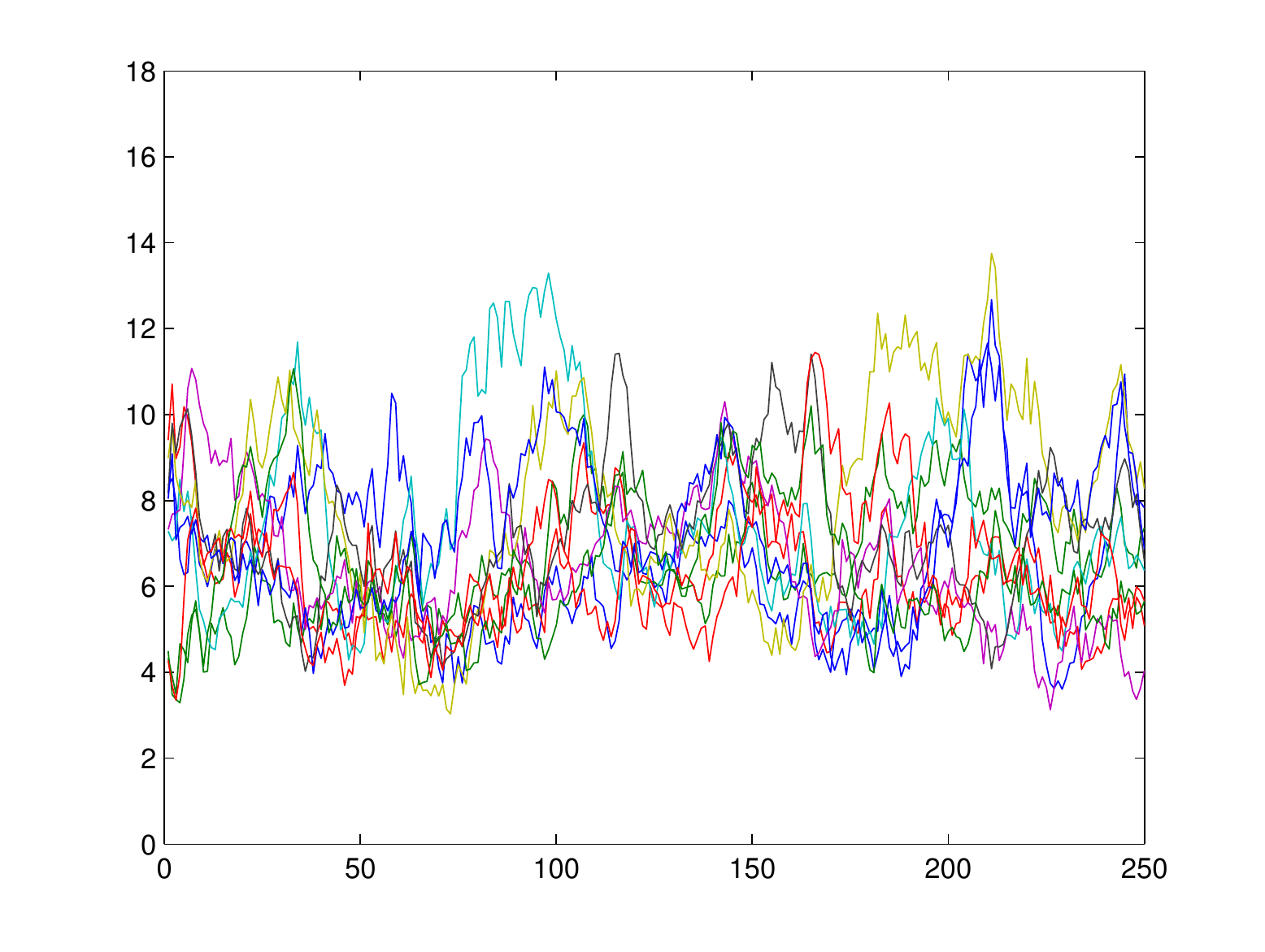}
                 \caption{Embedded HMM (0.81 seconds/sample)}
         \end{subfigure}
         \caption{Estimated autocorrelation times for each latent variable for Model 1, adjusted for computation time}
         \label{fig:acf}
\end{figure}

The computation time-adjusted autocorrelation estimates for Model 1, for all the latent variables, plotted over time, are presented in Figure \ref{fig:acf}. We found that the combination of single-state Metropolis and PGBS works best for the unimodal model. The other samplers work reasonably well too. We note that the spike in autocorrelation time for the PGBS and to a lesser extent for the PGBS with Metropolis sampler occurs at the point where the data is very informative. This in turn makes the use of the diffuse transition distribution the particles are drawn from inefficient and much of the sampling in that region is due to the Metropolis updates. Here, we also note that the computation time adjustment is sensitive to the particularities of the implementation, in this case done in MATLAB, where performance depends a lot on how well vectorization can be exploited. Implementing the samplers in a different language might change the relative comparisons.

We now look at how the samplers perform on the more challenging Model 2. We first did a preliminary check of whether the samplers do indeed explore the different modes of the distribution by looking at variables far apart in the sequence, where we expect to see four modes (with all possible combinations of signs). This is indeed the case for both the PGBS with Metropolis and Embedded HMM samplers.
\begin{figure}[t]
         \centering
         \begin{subfigure}[b]{0.45\textwidth}
                 \includegraphics[width=\textwidth]{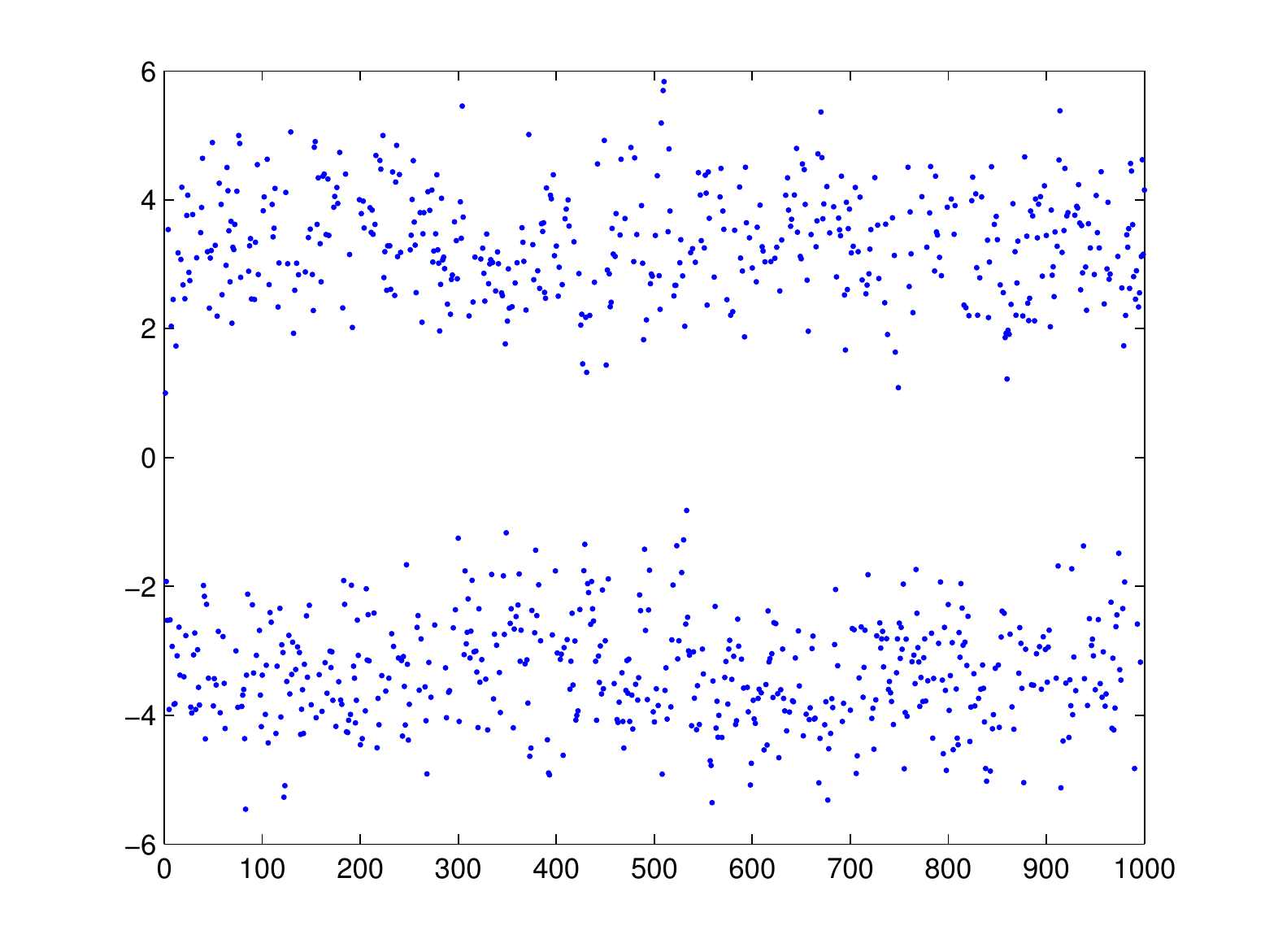}
                 \caption{Embedded HMM}
         \end{subfigure}
     ~
         \begin{subfigure}[b]{0.45\textwidth}
                 \includegraphics[width=\textwidth]{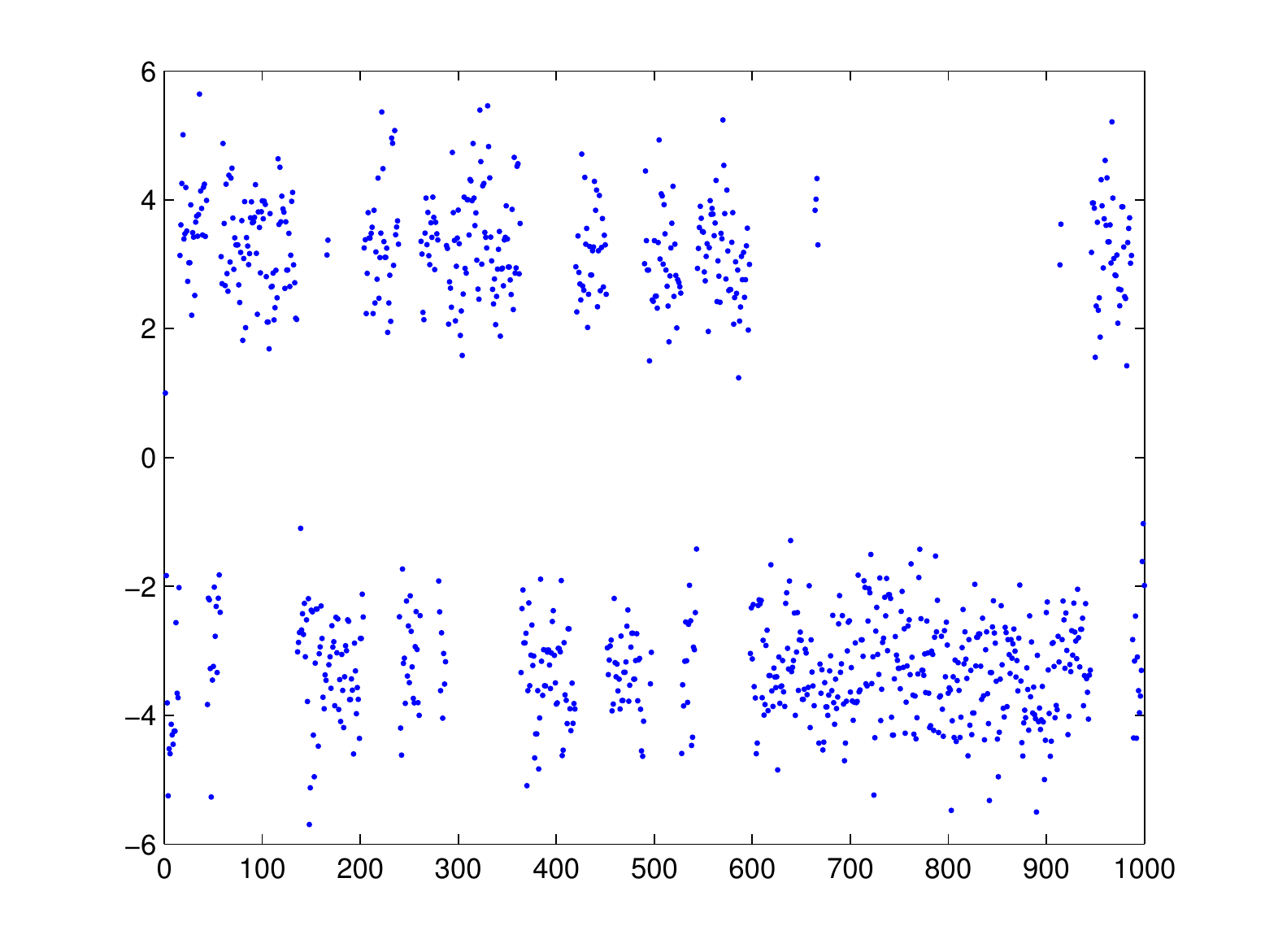}
                 \caption{Particle Gibbs with Metropolis}
         \end{subfigure}
         \caption{Comparison of Samplers for Model 2, $x_{1, 300}$}
         \label{fig:trace}
\end{figure}
Next, we look at how efficiently the latent variables are sampled. Of particular interest are the latent variables with well-separated modes, since sampling performance for such variables is illustrative of how well the samplers explore the different posterior modes. Consider the variable $x_{1, 300}$, which has true value $-1.99$. Figure \ref{fig:trace} shows how the different samplers explore the two modes for this variable, with equal computation times used to produced the samples for the trace plots. We can see that the embedded HMM sampler with flip updates performs significantly better for sampling a variable with well-separated modes. Experiments showed that the performance of the embedded HMM sampler on model 2 without flip updates is much worse.
\begin{figure}[t]
         \centering
         \begin{subfigure}[b]{0.45\textwidth}
                 \includegraphics[width=\textwidth]{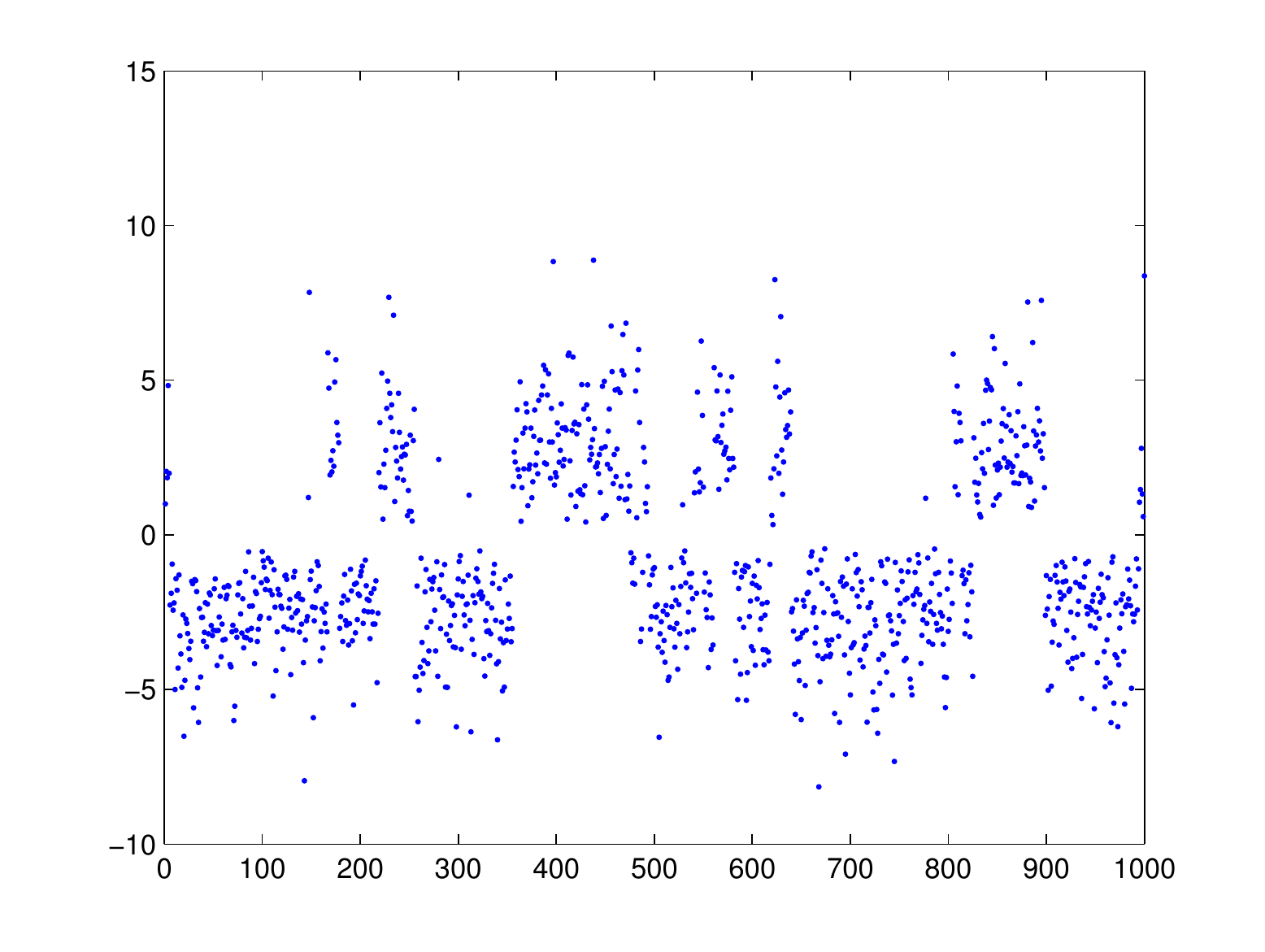}
                 \caption{Embedded HMM}
         \end{subfigure}
     ~
         \begin{subfigure}[b]{0.45\textwidth}
                 \includegraphics[width=\textwidth]{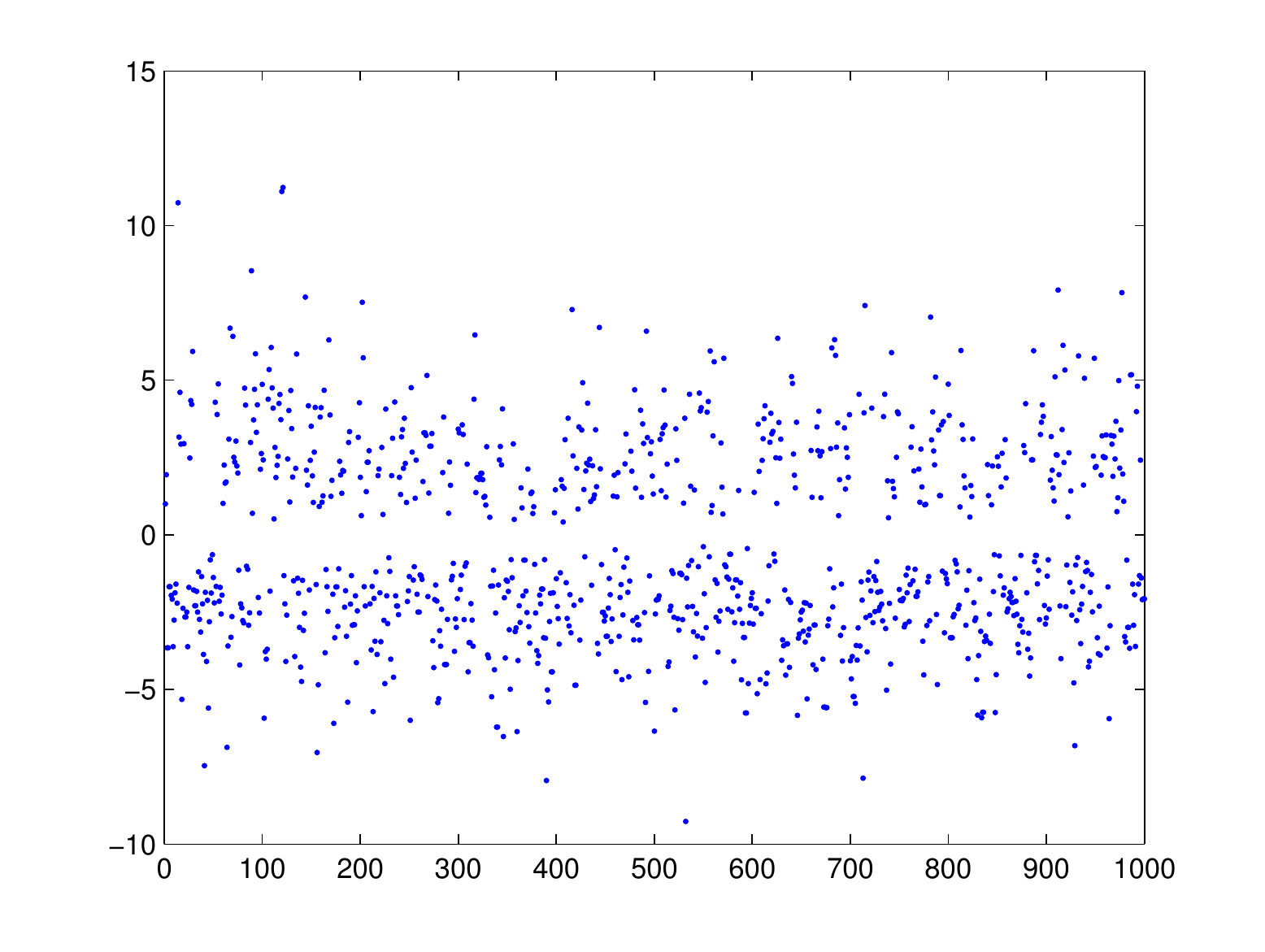}
                 \caption{Particle Gibbs with Metropolis}
         \end{subfigure}
         \caption{Comparison of Samplers for Model 2, $x_{3, 208}x_{4, 208}$}
         \label{fig:trace2}
\end{figure}

We can also look at the product of the two variables $x_{3, 208}x_{4, 208}$, with true value $-4.45$. The trace plot is given in Figure \ref{fig:trace2}. In this case, we can see that the PGBS with Metropolis sampler performs better. Since the flip updates change the signs of all dimensions of $x_{i}$ at once, we do not expect them to be as useful for improving sampling of this function of state. The vastly greater number of particles used by PGBS, $80000$, versus $80$ for the embedded HMM method, works to the advantage of PGBS, and explains the performance difference. 

\begin{figure}[t]
         \centering
         \begin{subfigure}[b]{0.45\textwidth}
                 \includegraphics[width=\textwidth]{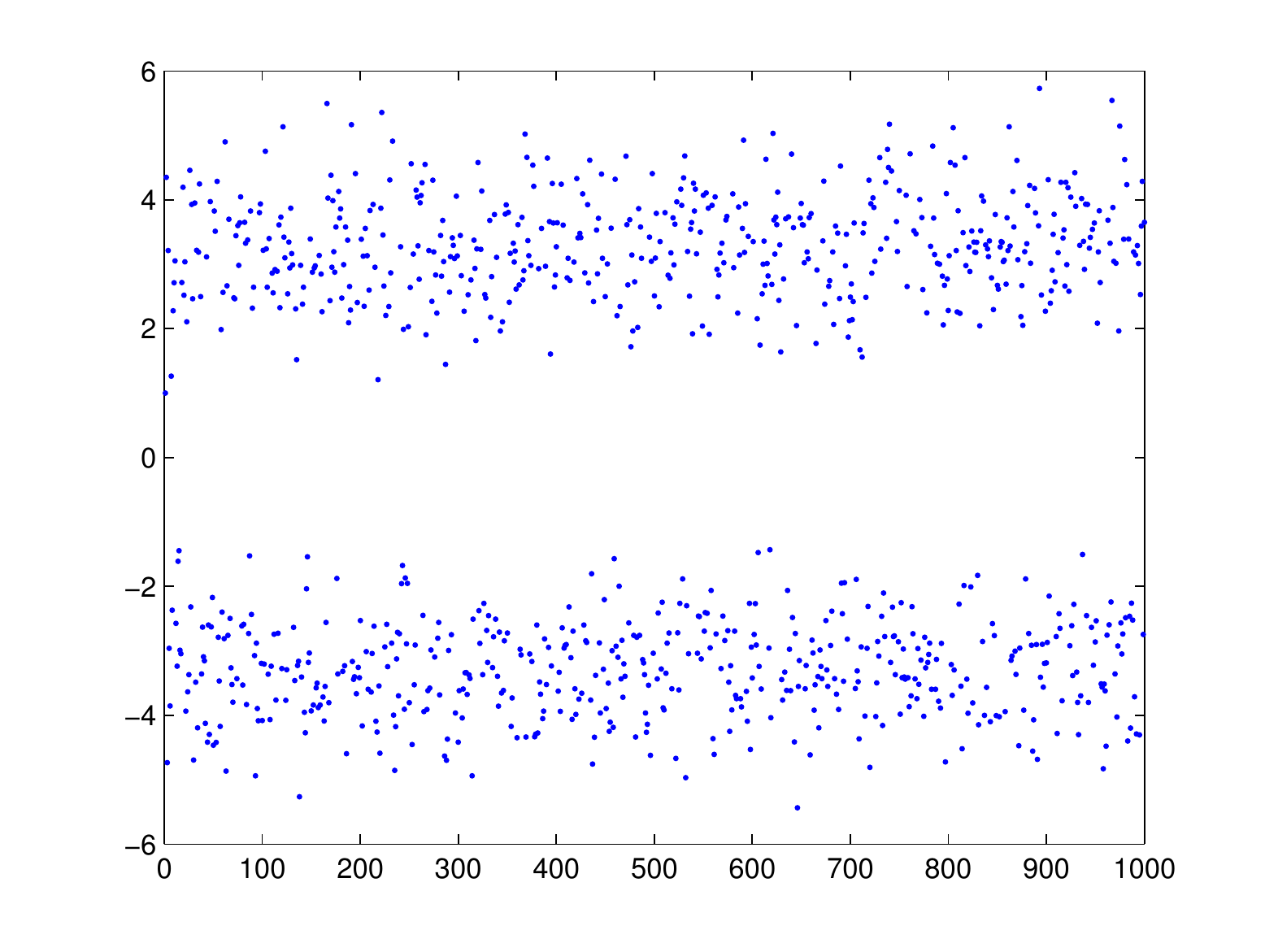}
                 \caption{Trace plot for $x_{1, 300}$}
         \end{subfigure}
     ~
         \begin{subfigure}[b]{0.45\textwidth}
                 \includegraphics[width=\textwidth]{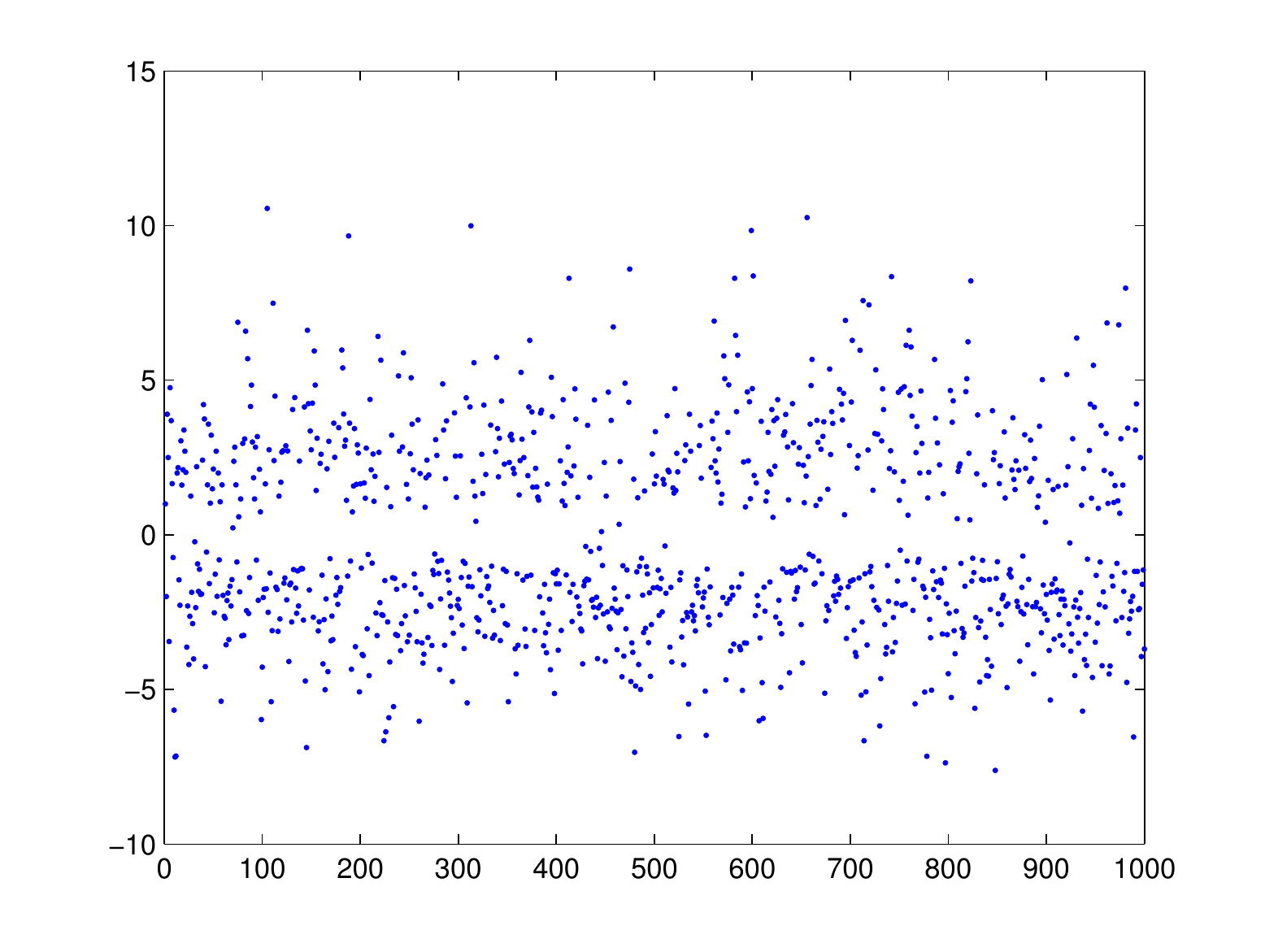}
                 \caption{Trace plot for $x_{3, 208}x_{4, 208}$}
         \end{subfigure}
         \caption{Combination of embedded HMM and PGBS with Metropolis samplers}
         \label{fig:trace3}
\end{figure}

Looking at these results, we might expect that we can get a good sampler for both $x_{1, 300}$ and $x_{3, 208}x_{4, 208}$ by alternating embedded HMM and PGBS with Metropolis updates. This is indeed the case, which can be seen in Figure \ref{fig:trace3}. For producing these plots, we used an embedded HMM sampler with the same settings as in the experiment for Model $2$ and a PGBS with Metropolis sampler with $10000$ particles and Metropolis updates using the same settings as in the experiment for Model $2$.

This example of Model 2 demonstrates another advantage of the embedded HMM viewpoint, which is that it allows us to design updates for sampling pool states to handle certain properties of the density. This is arguably easier than designing importance densities in high dimensions.

\section{Conclusion}

We have demonstrated that it is possible to use embedded HMM's to efficiently sample state sequences in models with higher dimensional state spaces. We have also shown how embedded HMMs can improve sampling efficiency in an example model with a multimodal posterior, by introducing a new pool state selection scheme. There are several directions in which this research can be further developed.

The most obvious extension is to treat the model parameters as unknown and add a step to sample parameters given a value of the latent state sequence. In the unknown parameter context, it would also be interesting to see how the proposed sequential pool state selection schemes can be used together with ensemble MCMC updates of Shestopaloff and Neal (2013). For example, one approach is to have the pool state distribution depend on the average of the current and proposed parameter values in an ensemble Metropolis update, as in Shestopaloff and Neal (2014).

One might also wonder whether it is possible to use the entire current state of $x$ in constructing the pool state density at a given time. It is not obvious how (or if it is possible) to overcome this limitation. For example, for the forward scheme, using the current value of the state sequence at some time $k > i$ to construct pool states at time $i$ means that the pool states at time $k$ will end up depending on the current value of $x_{k}$, which would lead to an invalid sampler.

At each time $i < n$, the pool state generation procedure does not depend on the data after time $i$, which may cause some difficulties in scaling this method further. On one hand, this allows for greater dispersion in the pool states than if we were to impose a constraint from the other direction as with the single-state Metropolis method, potentially allowing us to make larger moves. On the other hand, the removal of this constraint also means that the pool states can become too dispersed. In higher dimensions, one way in which this can be controlled is by using a Markov chain that samples pool states close to the current $x_{i}$ --- that is, a Markov chain that is deliberately slowed down in order not to overdisperse the pool states, which could lead to a collection of sequences with low posterior density.

\section*{Acknowledgements}

We thank Arnaud Doucet for helpful comments. This research was supported by the Natural Sciences and Engineering Research Council of Canada.  A.~S.\ is in part funded by an NSERC Postgraduate Scholarship. R.~N.\ holds a Canada Research Chair in Statistics and Machine Learning. 

\section*{References}

\begin{description}

\item
Andrieu, C., Doucet, A. and Holenstein, R. (2010) ``Particle Markov chain Monte Carlo methods'', {\em Journal of the Royal Statistical Society B}, vol.~72, pp.~269-342.

\item
Lindsten, F.; Schon, T.B. (2012) ``On the use of backward simulation in the particle Gibbs sampler'', in {\em Acoustics, Speech and Signal Processing (ICASSP), 2012 IEEE International Conference on}, pp.~3845-3848.

\item
Neal, R.M. (1998) ``Regression and classification using Gaussian process priors'', in J.M. Bernardo et al (editors) {\em Bayesian Statistics 6}, Oxford University Press, pp.~475-501.

\item
Neal, R. M. (2003) ``Markov Chain Sampling for Non-linear State Space Models using Embedded Hidden Markov Models'', Technical Report No. 0304, Department of Statistics, University of Toronto, http://arxiv.org/abs/math/0305039.

\item
Neal, R. M., Beal, M. J., and Roweis, S. T. (2004) ``Inferring state sequences for non-linear systems with embedded hidden Markov models'', in S. Thrun, et al (editors), {\em Advances in Neural Information Processing Systems 16}, MIT Press.

\item
Pitt, M. K. and Shephard, N. (1999) ``Filtering via Simulation: Auxiliary Particle Filters'', {\em Journal of the American Statistical Association}, vol.~94, no.~446, pp.~590-599.

\item
Shestopaloff, A. Y. and Neal, R. M. (2013) ``MCMC for non-linear state space models using ensembles of latent sequences'', Technical Report, http://arxiv.org/abs/1305.0320.

\item
Shestopaloff, A. Y. and Neal, R. M. (2014) ``Efficient Bayesian inference for stochastic volatility models with ensemble MCMC methods'', Technical Report, http://arxiv.org/abs/1412.3013.

\item Shestopaloff, A. Y. (2016) ``MCMC Methods for Non-Linear State Space Models''. PhD Thesis, University of Toronto.

\end{description}

\end{document}